\documentclass[conference]{IEEEtran}
\newcommand{\hpcayear}{2026}
\usepackage{cite}
\usepackage{amsmath,amssymb,amsfonts}
\usepackage{graphicx}
\usepackage{textcomp}
\usepackage{xcolor}
\usepackage[hyphens]{url}
\usepackage{fancyhdr}
\usepackage{stmaryrd}
\usepackage{listings}
\usepackage{bm}
\usepackage{subcaption}

\usepackage{algorithm2e}
\SetKwComment{Comment}{/* }{ */}
\SetKwInOut{Input}{Input}
\RestyleAlgo{ruled}

\usepackage{braket}
\usepackage{float}

\usepackage{lipsum}
\def\BibTeX{{\rm B\kern-.05em{\sc i\kern-.025em b}\kern-.08em
    T\kern-.1667em\lower.7ex\hbox{E}\kern-.125emX}}

\begin{document}

\title{Fully Parallelized BP Decoding for Quantum LDPC Codes Can Outperform BP-OSD}

\author{\IEEEauthorblockN{Ming Wang}
\IEEEauthorblockA{\textit{Department of Computer Science} \\
\textit{North Carolina State University}\\
Raleigh, USA \\
mwang42@ncsu.edu}
\and
\IEEEauthorblockN{Ang Li}
\IEEEauthorblockA{\textit{Pacific Northwest National Laboratory}\\
Richland, Washington \\
ang.li@pnnl.gov}
\and
\IEEEauthorblockN{Frank Mueller}
\IEEEauthorblockA{\textit{Department of Computer Science} \\
\textit{North Carolina State University}\\
Raleigh, USA \\
fmuelle@ncsu.edu}}
\def\aeopen{}           
\def\aereviewed{}     
\def\aereproduced{} 
\fancypagestyle{camerareadyfirstpage}{%
  \fancyhead{}
  \renewcommand{\headrulewidth}{0pt}
  \fancyhead[C]{
    \ifdefined\aeopen
    \parbox[][12mm][t]{13.5cm}{\hpcayear{} IEEE International Symposium on High-Performance Computer Architecture (HPCA)}    
    \else
      \ifdefined\aereviewed
      \parbox[][12mm][t]{13.5cm}{\hpcayear{} IEEE International Symposium on High-Performance Computer Architecture (HPCA)}
      \else
      \ifdefined\aereproduced
      \parbox[][12mm][t]{13.5cm}{\hpcayear{} IEEE International Symposium on High-Performance Computer Architecture (HPCA)}
      \else
      \parbox[][0mm][t]{13.5cm}{\hpcayear{} IEEE International Symposium on High-Performance Computer Architecture (HPCA)}
    \fi 
    \fi 
    \fi 
    \ifdefined\aeopen 
      \includegraphics[width=12mm,height=12mm]{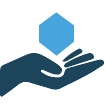}
    \fi 
    \ifdefined\aereviewed
      \includegraphics[width=12mm,height=12mm]{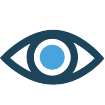}
    \fi 
    \ifdefined\aereproduced
      \includegraphics[width=12mm,height=12mm]{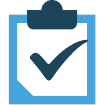}
    \fi
  }
  \fancyfoot[C]{}
}
\maketitle
\begin{abstract}
This work presents a hardware-efficient and fully parallelizable
decoder for quantum LDPC codes that leverages belief propagation (BP)
with a speculative post-processing strategy inspired by classical
Chase decoding algorithm. By monitoring bit-level oscillation patterns
during BP, our method identifies unreliable bits and generates
multiple candidate vectors to selectively flip syndromes. Each
modified syndrome is then decoded independently using short-depth BP,
a process we refer to as BP-SF (syndrome flip).  This design
eliminates the need for costly Gaussian elimination used in the
current BP-OSD approaches. Our implementation achieves logical error
rates comparable to or better than BP-OSD while offering significantly
lower latency due to its high degree of parallelism for a variety of
bivariate bicycle codes. Evaluation on the
$\llbracket 144,12,12\rrbracket$ bivariate bicycle code shows that the
proposed decoder reduces average latency to approximately $70\%$ of
BP-OSD. When post-processing is parallelized the average latency is
reduced to $55\%$ compared  to the single process
implementation, with the maximum latency reaching as low as $18\%$.  These
advantages make it particularly well-suited for real-time and
resource-constrained quantum error correction systems.

\end{abstract}
{\renewcommand{\thefootnote}{}\footnotetext{An earlier version of this paper was published in the proceedings of the IEEE International Symposium on High-Performance Computer Architecture (HPCA) 2026. This version corrects typographical and grammatical errors.}}

\section{Introduction}
One of the central challenges in quantum computing is achieving
fault-tolerant quantum computation (FTQC), which requires the use of
quantum error correction (QEC) codes to correct errors arising from
noisy quantum devices.  Significant efforts~\cite{9773217, 10764437,
  10.1145/3579371.3589037, 10.1145/3503222.3507707,
  10.1145/3695053.3731022} have been made toward designing decoders
and hardware architectures for various QEC codes. Among the QEC codes,
quantum low-density parity-check (qLDPC) codes have attracted
significant attention in recent years due to their potential to encode
more logical qubits than surface codes while maintaining a high
threshold~\cite{hgp,panteleev2021degenerate,Bravyi_2024}. However,
decoding qLDPC codes remains a major challenge. To ensure reliable
operation, especially in fault-tolerant quantum memory and
computation, errors must be corrected both quickly and accurately.  In
practice, decoders must keep pace with the rate of syndrome extraction
to prevent data backlog~\cite{RevModPhys.87.307}, placing stringent
demands on both the decoding algorithms and their implementations in
terms of performance and computational efficiency.

Belief propagation (BP)-based decoders, widely used in classical LDPC
codes, are appealing due to their low complexity, parallelizability,
and near-optimal performance~\cite{910578, 1657127, su202258, 5437474}. 
However, their effectiveness diminishes
significantly when applied to qLDPC codes. This degradation is
primarily due to inherent properties of qLDPC codes, such as
degeneracy, the presence of many low-weight stabilizers, and the
prevalence of trapping sets, which hinder
convergence and reliability of BP decoding~\cite{raveendran2021trapping}.

Many works have sought to address the limitations of BP-based decoders
on qLDPC codes. In~\cite{poulin2008iterative}, Poulin and Chung
proposed several techniques to enhance convergence, including random
freezing of variable nodes, perturbing prior information, and
colliding unsatisfied check nodes. Similar approaches are also adopted
in~\cite{10619083} to freeze a node based on posterior
information. In~\cite{10409156}, instead of performing static trapping
set analysis prior to decoding, Chytas et al.
identified oscillation bits affected by trapping sets dynamically
during the decoding process. Once identified, the posterior
information of these bits is modified, similar to the approach
in~\cite{poulin2008iterative}, to help the decoder escape local minima
and converge. Raveendran et al.~\cite{raveendran2021trapping} analyzed
different types of trapping sets and proposed using a layered BP
decoder to mitigate the effect of symmetric trapping sets. While
layered decoder can reduce complexity, it often comes at the cost of
increased decoding latency as layered decoders are serial. More
recently, Yin et al.~\cite{yin2024symbreak} leveraged the degeneracy
property of quantum LDPC codes to enhance BP decoding. By analyzing
bit-wise marginal probabilities from BP, they selectively split rows
in the parity-check matrix and modify the corresponding Tanner graph.
This symmetry-breaking technique helps the decoder avoid convergence
stalls caused by structural degeneracies in the code. In general,
these methods typically involve modifications to the graph structure,
prior, or posterior information in BP decoding to constrain the
decoder, which effectively reduce the search space to facilitate
convergence.
In~\cite{gong2024toward}, Gong et al. proposed guided decimation
guessing (GDG), a method based on tracking the decoding history of BP
to accelerate its convergence. They also employ a tree-search-like
strategy that keeps multiple decimation paths open to correct errors
as a decoding ensemble. However, the decision tree-like structure of
the ``guessing'' phase restricts the algorithm's potential for further
parallelization. During the preparation of this work, M{\"u}ller et
al.~\cite{muller2025improved} proposed Relay-BP, a hardware-friendly
and real-time decoder constructed by chaining multiple
Mem-BP~\cite{11027786} decoders with varying memory strengths. While
this approach achieves a logical error rate significantly lower than
the BP-OSD decoder, its sequential chaining structure imposes a
latency overhead that prevents the different decoding stages from
being parallelized.

\begin{figure*}[!h]
    \centering
    \includegraphics[width=5.4 in]{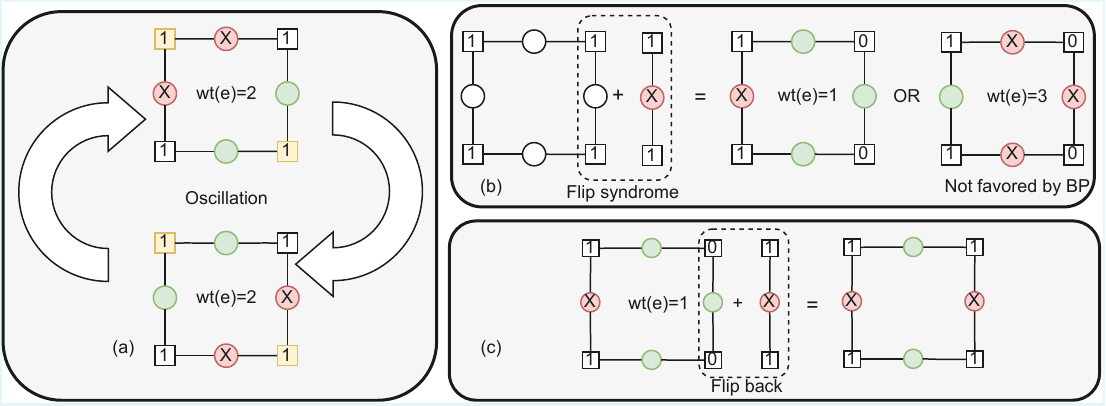}
    \caption{(a) Example of BP failing to converge due to
      oscillations. Red Xs denote bits identified as erroneous by BP,
      and yellow squares represent unsatisfied syndrome checks. All
      four bits are oscillating.  (b) One oscillating bit (e.g., the
      rightmost one) is selected, and its neighboring syndromes are
      flipped. BP then converges since the two competing error
      patterns now have different weights.  (c) After convergence, the
      selected bit is flipped back to restore consistency with the
      original input syndrome.}
    \label{scheme}
\end{figure*}

Apart from efforts aimed at improving the BP decoder itself, several
works focus on post-processing techniques to enhance decoding
performance. A widely used approach is belief propagation combined
with ordered statistics decoding
(BP-OSD)~\cite{panteleev2021degenerate, roffe2020}. While BP-OSD
significantly enhances error correction performance, its reliance on a
Gaussian elimination step during the OSD phase introduces substantial
computational overhead. Specifically, this step incurs a complexity of
$O(N^3)$~\cite{panteleev2021degenerate}, where $N$ denotes either the
code length (in the code-capacity error model) or the number of error
mechanisms (in the circuit-level noise model).
In contrast, each iteration of BP has a computational complexity of
only $O(N)$~\cite{910578}. This difference makes BP-OSD
computationally expensive and less suitable for large-scale or
real-time decoding applications. To address this limitation, recent
works such as~\cite{wolanski2024ambiguity, hillmann2024localized}
proposed
partitioning the Tanner graph into several clusters. This localized
decoding approach reduces the size of the matrices involved in
Gaussian elimination, thereby lowering the overall computational
burden without substantially compromising decoding performance.  These
post-processing techniques provide a fallback when BP decoding fails,
but they often involve complex data structures and control flows,
making them difficult to implement efficiently in hardware.

In this work, rather than focusing on improving BP or OSD
individually, we propose that we can achieve error rate performance
comparable to BP-OSD while offering significantly lower latency. The
key insights behind our approach, which we refer to as BP-SF, are
twofold: (i) Most BP decoding failures are caused by a small subset of
bits referred to as oscillating bits. By flipping some of these bits
in the syndrome domain, BP can often converge rapidly. (ii) Due to the
low complexity and inherently parallel nature of BP, we can
speculatively perform multiple decoding attempts in parallel,
incurring minimal additional latency. Our technique can also be viewed
as analogous to providing different starting points to the optimizer,
helping it escape local minima and find a global one.

We give a brief visualization in Figure~\ref{scheme} to illustrate our
purely BP-based algorithm. We demonstrate that, with appropriate
implementation and design, it can closely match the logical error
rates achieved by BP-OSD. Through simulations on a variety of quantum
LDPC codes, including the $\llbracket 72,12,6 \rrbracket$,
$\llbracket 144,12,12 \rrbracket$, and
$\llbracket 288,12,18 \rrbracket$ bivariate bicycle codes; the
$\llbracket 126,12,10 \rrbracket$ and $\llbracket 154,6,16 \rrbracket$
coprime bivariate bicycle codes; the $\llbracket 225,16,8 \rrbracket$
subsystem hypergraph product simplex code and the
$\llbracket 254,28 \rrbracket$ generalized bicycle code,
we show that the proposed decoder performs similarly to BP-OSD with a
combination-sweep of order 10. Additionally, the proposed BP decoder
requires lower latency as it is fully parallelizable, which means it
can outperform BP-OSD in execution efficiency, making it a promising
candidate for scalable and efficient decoding of quantum LDPC codes in
practical fault-tolerant quantum computing systems.

\textbf{Contributions:}
\begin{itemize}
\item \textbf{BP Decoding Analysis:} We analyze the behavior of the BP
  decoding algorithm using the $\llbracket 144,12,12 \rrbracket$ code
  as a case study, revealing a long-tail distribution in the number of
  iterations and a strong correlation between oscillating bits and
  actual error locations.
    
\item \textbf{Oscillation-Guided Speculative Decoding:} We propose a
  fully parallelizable post-processing technique that leverages
  bit-level oscillation during BP decoding to identify unreliable bits
  and generate test vectors, enabling Chase-like decoding without
  requiring Gaussian elimination.
    
\item \textbf{Efficient Implementation:} We implement the proposed
  decoder in both serial and multi-process CPU versions, achieving
  faster runtime than BP-OSD while maintaining comparable logical
  error rates across various qLDPC codes. We also explore GPU
  implementation potential, and provide a pessimistic upper bound for
  GPU decoding time.
\end{itemize}

\section{Background}
\subsection{Quantum LDPC Codes}

Stabilizer codes are among the most commonly used codes in quantum
error correction. One can measure each stabilizer to infer both the
type and location of errors in a multi-qubit system. To construct such
a code, all stabilizers must commute with each other. Thus, they have
a common eigenspace and form a stabilizer group $\mathcal{S}$. The
code space $\mathcal{C}$ defined by such group is
\begin{equation}
    \mathcal{C}=\{\ket{\psi} |\: s\ket{\psi}=\ket{\psi},\: \forall s\in \mathcal{S}\}.
\end{equation}

An $\llbracket n, k, d\rrbracket $ stabilizer code can be defined by
$n-k$ independent stabilizers, allowing us to encode $k$ qubits of
logical information into an $n$-qubit block tolerating up to
${\lfloor (d-1)/2 \rfloor}$ errors. CSS codes are an important class
of stabilizer codes with two sets of stabilizers, $X$-type and $Z$-type,
represented by parity-check matrices $H_X$ and $H_Z$,
respectively. Each row in a parity-check matrix corresponds to a
stabilizer generator, and each column corresponds to a physical
qubit. A ``1'' entry indicates an $X$ or $Z$ operator (depending on
whether it is in $H_X$ or $H_Z$), while a ``0'' indicates the
identity. Consequently, an $X$-type stabilizer acts as $X$ or the
identity on each qubit, and a $Z$-type stabilizer acts as $Z$ or the
identity on each qubit. Errors can therefore be corrected by handling
$Z$ errors and $X$ errors separately. Since all stabilizers must
commute with each other, it follows directly that for a CSS code
$H_XH_Z^T=0$. If both $H_X$ and $H_Z$ are sparse matrices, the code is
a CSS-type qLDPC code. The sparsity of these matrices offers a key
advantage: syndrome extraction can be performed using fewer quantum
gates, thereby reducing circuit depth and potential error
accumulation. In this sense, the surface code can be regarded as a
special case of qLDPC codes, characterized by strictly local,
nearest-neighbor interactions on a 2D lattice. More general qLDPC
codes, in contrast, typically exhibit higher connectivity and
non-local parity-check relationships. This allows them to encode more
logical qubits while maintaining a large code distance, but it also
makes matching-based algorithms (commonly used for surface codes) less
effective, since the increased connectivity introduces hyperedges into
the decoding graph.

\subsection{Decoding Problem}
Assuming the noise is uniform on each bit, the optimal syndrome
decoding problem for classical codes can be formalized as follows:
Given a code with parity-check matrix $H\in\mathbb{F}_2^{M\times N}$
and syndrome $\bm{s}\in \mathbb{F}_2^M$, we want to find an error
$\hat{\bm{e}}\in\mathbb{F}_2^N$ that satisfies the syndrome
\begin{equation}
\label{MLD_classical}
    \hat{\bm{e}}=\arg\min_{\hat{\bm{e}}H^T=\bm{s}}(\sum_i^N \hat{\bm{e}_i}).
\end{equation}

Quantum stabilizer codes face a problem due to the phenomenon of
\emph{degeneracy}, where multiple errors have the same effect on the
code space. As a result, the goal of decoding is not to identify the
most likely error itself, but rather the most likely equivalence class
of errors modulo stabilizers, since errors that differ by a stabilizer
operation act identically on the code space.
Given a syndrome, $\bm{s}$, the optimal decoding problem considering
degeneracy becomes
\begin{equation}
\label{MLD_quantum}
    [\hat{E}] = \arg\max_{[E]: \text{synd}(E) = \bm{s}} \Pr([E]),
\end{equation}
where $[E]$ denotes the equivalence class of errors under the
stabilizer group and $\Pr([E])$ is the total probability of all errors
in that class. While the classical decoding problem in Eq.~(\ref{MLD_classical}) is
NP-hard~\cite{1055873}, its quantum counterpart in Eq.~(\ref{MLD_quantum}) is even
more complex, being \#P-complete~\cite{10.1109/TIT.2015.2422294}. In practice, optimal decoding is
computationally intractable, so efficient decoders typically aim to
approximate the solution to Eq.~(\ref{MLD_classical}) with good
performance under realistic noise models.

As CSS codes can be decoded separately on the $X$- and $Z$-error
bases, each decoding problem can be treated as a classical decoding
task and addressed using classical decoding algorithms. For example,
BP, the most commonly used decoder for classical LDPC codes, is also
widely applied in qLDPC codes. This is because the sparsity of qLDPC
and LDPC codes can be effectively exploited by BP-based decoders,
which rely on the assumption of independent probability updates. But
this assumption is only
valid when the parity-check matrix is sparse. As the matrix becomes
denser, the variable nodes exhibit stronger correlations, violating
the independence assumption. Consequently, although the BP decoder
can, in principle, be applied to denser codes, its empirical
performance tends to degrade due to the increased dependency within
the graph structure.
Given an $M \times N$ parity-check matrix $H$, let $v_1, \dots, v_N$
denote the variable nodes (corresponding to the columns of $H$) and
$c_1, \dots, c_M$ the check nodes (corresponding to the rows). The
normalized min-sum algorithm, a widely used variant of BP, can be
described as follows:

\begin{enumerate}

\item \textbf{Initialization:} Given the prior error information,
  $\bm{p}$, of each variable node, $v_i$, the channel LLR is
  initialized to
  \begin{equation}
    l^{ch}_{v_i}=\log \frac{1-p_{v_i}}{p_{v_i}}.
  \end{equation}

\item \textbf{Variable-to-Check (V2C) Message Update:} Each variable
  node, $v_i$, sends a message to its neighboring check node, $c_j$,
  based on the channel LLR and the incoming messages from all other
  neighboring check nodes, denoted as
  \begin{equation}
    l_{v_i\rightarrow c_j} = l^{ch}_{v_i}+\sum_{c_{j'}\in N(v_i) \setminus \{c_j\}} l_{c_{j'}\rightarrow v_i}\:\:,
  \end{equation}
  where $l_{c_{j'}\rightarrow v_i}$ is set to 0 for the first
  iteration, ``$\setminus$'' is set minus, and $N(v_i)$ is the set of
  all the check nodes connected with $v_i$.

\item \textbf{Check nodes to variable nodes (C2V) update:} Each check
  node, $c_j$, updates its message to a neighboring variable node,
  $v_i$, using the min-sum rule denoted as
  \begin{equation}
  \label{c2v}
          \begin{split}
    l_{c_j\rightarrow v_i} = &(-1)^{s_j}\cdot \alpha\min_{v_{i'}\in N(c_j)\setminus\{v_i\}}|l_{v_{i'}\rightarrow c_j}|\cdot\\ &\prod_{v_{i'}\in N(c_j)\setminus\{v_i\}}\text{sign}(l_{v_{i'}\rightarrow c_j}),
    \end{split}
  \end{equation}

  where $\alpha$ is the damping factor used to attenuate the c2v
  message.
    
\item \textbf{Hard decision}: After a fixed number of iterations or
  upon convergence, the final marginal LLR for each variable node is
  computed as
  \begin{equation}
    l^{out}_{v_i} = l^{ch}_{v_i} +\sum_{c_{j'}\in N(v_i)} l_{c_{j'}\rightarrow v_i}\:\:,
  \end{equation}
  where the $\bm{l}^{out}$ is the marginalized LLR for each variable
  node. The estimated error is then obtained via hard decision as
  \begin{equation}
    \hat{e_i} = \begin{cases}
      0& \text{if} \quad l_{v_i}^{out}>0\\
      1& \text{otherwise}
    \end{cases},
  \end{equation}
\end{enumerate}
  
In each iteration, steps 2, 3 and 4 are performed. The BP decoding
algorithm proceeds until $\hat{\bm{e}}H^T=\bm{s}$ is satisfied or the
maximum number of iterations is reached.

\section{BP Behavior Analysis: A Case Study}

In this section, we analyze the behavior of BP-based decoders on qLDPC
codes using the $\llbracket144,12,12\rrbracket$ ``gross'' code
from~\cite{Bravyi_2024} as a representative case study. While specific
to this code, the analysis provides general insights into the behavior
of BP decoding on qLDPC codes and offers guidance for improving
decoder performance. The BP decoder used below is a min-sum decoder as
described in Eq.~(\ref{c2v}) with an adaptive damping factor of
$\alpha=1-2^{-i}$, where $i$ is the current number of iterations.

\subsection{Number of Iterations}

\begin{figure}[htb]
    \centering
    \includegraphics[width=3.0 in]{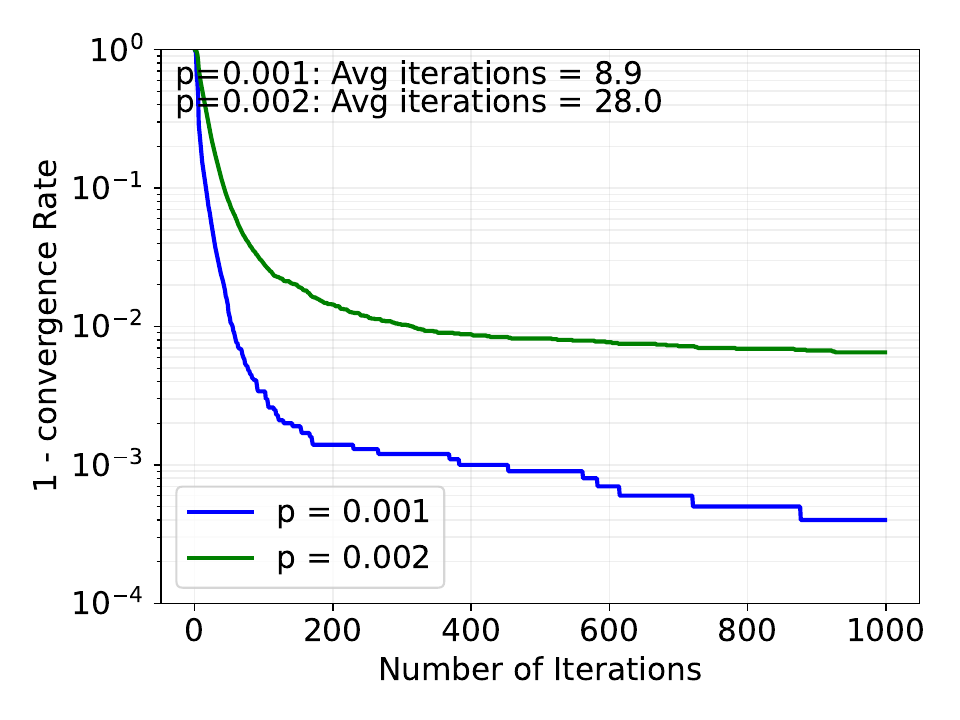}
    \caption{Ratio of unsuccessful BP decoding ($1-$convergence rate) on the
      $\llbracket144,12,12\rrbracket$ code under the circuit-level
      noise model. The maximum number of decoder iterations is set to 1,000
      and number of samples is 10,000. 
        }
    \label{idist}
\end{figure}

Figure~\ref{idist} shows the ratio of syndromes failed to converge 
under the circuit-level noise model, where $p$ denotes the
physical error rate. The curves are obtained by simulating 10{,}000
samples for each of $p = 0.001$ and $p = 0.002$, both representative
values below the threshold. For each sample, we record the
number of iterations required for convergence and compute the
cumulative distribution. This non-convergence rate at iteration $i$ is
defined as the fraction of samples that have not converged within $i$
iterations—that is, $1$ minus the cumulative convergence rate.

As Fig.~\ref{idist} indicates, in most cases, BP converges within a
small number of iterations. For instance, at $p=0.001$, the average
number of iterations is merely 8.9 despite setting maximum number of
iterations to 1,000.  Even at higher error rates, such as $p=0.002$,
the average number of iterations remains low, although the tail
becomes longer. Notably, cases that do not converge within the early
iterations rarely benefit from increasing the iteration count
further. This observation motivates an alternative strategy: Rather
than extending the number of BP iterations, we can vary the inputs to
the BP decoder while keeping the maximum number of iterations
small. If these varied inputs have better independent chances of
successful decoding, then running multiple instances in parallel
allows us to exponentially suppress the logical error rate without
incurring significant decoding latency.

\subsection{Oscillation}

\begin{figure}[htb]
    \centering
    \includegraphics[width=3.0 in]{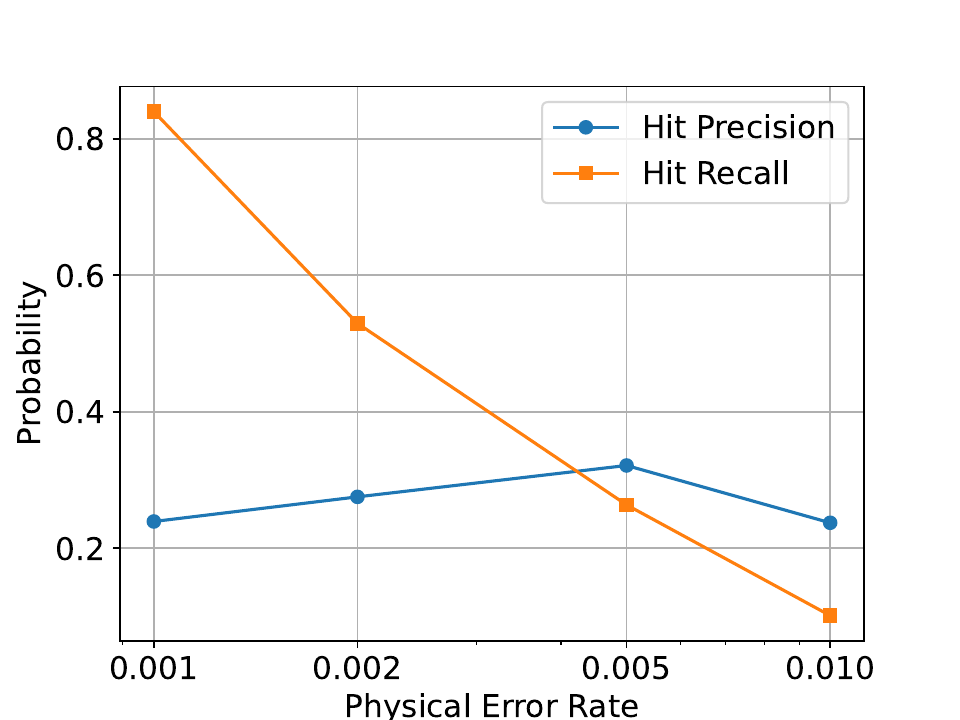}
    \caption{Precision and recall probabilities of candidate bit
      selection on the $\llbracket144,12,12\rrbracket$ code. We
      evaluate the correlation between candidate bits and actual error
      locations by identifying the top 50 most frequently flipped bits
      among approximately 8,000 error mechanisms. The decoder is run
      with a maximum of 50 iterations, and statistics are collected
      over 1,000 decoding failures. This analysis reveals how well
      bit-level oscillation can serve as a heuristic for error
      localization.
    }
    \label{osc}
\end{figure}

As suggested in previous works~\cite{raveendran2021trapping,
  10409156}, trapping sets and code degeneracy often result in
ambiguous BP decoding, leading to convergence failures. A common
symptom of such a failure is bit-level oscillation, where certain
output bits repeatedly flip between 0 and 1 across iterations. To
better understand the relationship between oscillating bits and
decoding errors, we analyze the dynamics of bit oscillations during
the BP process. During the decoding process, we track bit-level
oscillations by comparing the output of each iteration with that of
the previous one and counting how often each bit flips, which is
similar to~\cite{10409156}. We then identify a set of oscillating
bits, denoted by $\Phi$, based on their flip frequency. Specifically,
$\Phi$ is defined as the top $|\Phi|$ of the most frequently flipped
bits. We denote $\operatorname{supp}(\bm{e})$ as the set of erroneous
bits and define the hit precision and recall as
\begin{align}
  \text{Precision} &= \frac{|\operatorname{supp}(\bm{e}) \cap \Phi|}{|\Phi|},  \\
  \text{Recall} &= \frac{|\operatorname{supp}(\bm{e}) \cap \Phi|}{|\operatorname{supp}(\bm{e})|}.
\end{align}
Fig.~\ref{osc} shows the precision and recall rate when the min-sum
decoder fails to decode a syndrome. We can see that even when the BP
decoder fails to fully correct an error, the pattern of bit
oscillations essentially reveals a meaningful subset of the actual
error locations. In particular, at lower physical error rates, the set
of oscillating bits nearly covers the entire true error positions. To
confirm, we observe that the hit precision, i.e., the fraction of
oscillating bits that are indeed erroneous, is substantially higher
than the physical error rate.  This suggests that bits in the
oscillation set $\Phi$ are much more likely to be true errors than
random guesses, making them valuable targets for post-processing. As
the physical error rate increases, the recall decreases, primarily
because the total number of errors grows while the candidate set size
remains fixed.

\section{A New Speculative Decoding Method}

\begin{figure}[htb]
    \centering
    \includegraphics[width=2.2 in]{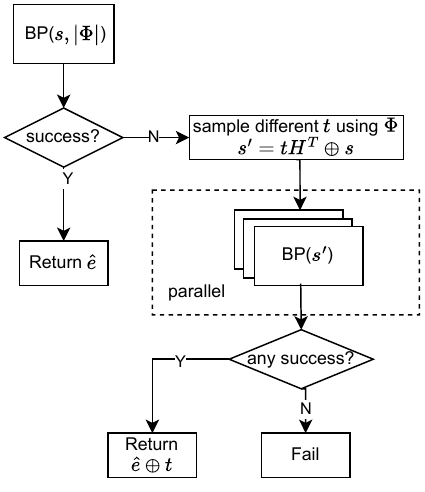}
    \caption{A simplified flowchart of the proposed decoder. The full
      procedure is described in Algorithm~\ref{alg:mybp}.}
    \label{flow}
\end{figure}

To enhance the performance of BP decoding, we adopt a Chase-like
post-processing technique~\cite{1054746} that we call BP-SF (Syndrome
Flip). This approach generates a set of trial vectors by flipping
candidate bits and attempts decoding on each of them, thereby
increasing the likelihood of successful error correction. In the
quantum decoding setting, we only have access to the syndrome and
prior information of error mechanisms, rather than the prior
information of each bit based on signal strength in classical soft
decoding. Therefore, unlike the original Chase algorithm, which relies
on prior channel information to select candidate bits, our BP-SF
method identifies candidate bits based on BP flipping statistics. As
shown in Figure~\ref{flow}, once these candidate bits $\Phi$ are
identified, we generate diverse decoding attempts by flipping the
input syndrome accordingly across multiple BP instances. This strategy
increases the variety in the decoder's inputs and distinguishes our
BP-SF approach from that in~\cite{koutsioumpas2025automorphism}, which
modifies the posterior information instead of the syndrome. If the
decoder successfully converges on this modified input, we then flip
these bits back in the output. This restoration ensures that the final
output error matches the original syndrome and preserves the validity
of the decoding result. Since these candidate bits are likely to
correspond to actual error locations, flipping them can also
effectively reduce the number of errors in the input and equivalently
lower the physical error rate. This reduction not only increases the
likelihood of successful decoding but also reduces the number of
iterations needed for BP to converge. The detailed pseudo code can be
found in Algorithm~\ref{alg:mybp}.
\begin{algorithm}[htb]
\caption{BP decoding with our Chase-like post-processing (BP-SF)}
\label{alg:mybp}
\SetKwInOut{Input}{Input}
\Input{Syndrome $\bm{s}$, max flip weight $w_{\max}$, number of uncertain bits $|\Phi|$}
\KwResult{Estimated error $\hat{\bm{e}}$}
\SetKwFor{ForPar}{parallel for}{}{end}

\SetKwFunction{FMain}{Main}
\SetKwFunction{FBP}{BP\_with\_oscillation}
\SetKwProg{Fn}{Function}{:}{}
\Fn{\FMain{$\bm{s}, w_{\max}, |\Phi|$}}{
    \Comment{Initial BP attempt with oscillation tracking} 
    $s_{\text{ucc}}, \hat{\bm{e}}, \Phi \gets \FuncSty{BP\_with\_oscillation}(\bm{s}, |\Phi|)$\\
    
    \uIf{$s_{\text{ucc}}$}{
        \Return $\hat{\bm{e}}$ \Comment{Syndrome decoded successfully}
    }
    
    \Else{
        \Comment{Speculative decoding using trial vectors based on $\Phi$} 
        \ForPar{$\bm{t} \in \FuncSty{combinations}(\Phi, w_{\max})$}{
            $\bm{s}' = \bm{s} \oplus \bm{t} H^T$ \Comment{Flip selected bits in syndrome domain}
            $s_{\text{ucc}}, \hat{\bm{e}} \gets \FuncSty{BP}(\bm{s}')$\\
            \uIf{$s_{\text{ucc}}$}{
                \Return $\hat{\bm{e}} \oplus \bm{t}$ \Comment{Undo flipped bits in output}
            }
        }
        \Return Decoding failure
    }
}
\vspace{0.3em}
\SetKwProg{Pn}{Function}{:}{}
\Pn{\FBP{$\bm{s}, |\Phi|$}}{
    $\text{flip\_count} \gets \bm{0}$\\
    $\hat{\bm{e}}_{\text{prev}} \gets \bm{0}$\\
    
    \For{$i = 1$ \KwTo $i_{\max}$}{
        $\hat{\bm{e}} \gets \FuncSty{BP\_Update}()$ \Comment{Standard BP update}
        $\text{flip\_count} \gets \text{flip\_count} + (\hat{\bm{e}} \oplus \hat{\bm{e}}_{\text{prev}})$ \Comment{Track bit oscillations}
        $\hat{\bm{e}}_{\text{prev}} \gets \hat{\bm{e}}$\\
        \uIf{$\hat{\bm{e}} H^T = \bm{s}$}{
            \Return \texttt{True}, $\hat{\bm{e}}, \emptyset$
        }
    }
    \Comment{Select top $|\Phi|$ most frequently flipped bits}
    $\Phi \gets \FuncSty{top}(\text{flip\_count}, |\Phi|)$\\
    \Return \texttt{False}, $\hat{\bm{e}}, \Phi$
}
\end{algorithm}

We note that this concept is also widely adopted in classical decoding
frameworks~\cite{1054746, 4036366, 10.1109/TCOMM.2009.07.070621,
  WANG2023102194}. In such approaches, multiple modified inputs are
generated and decoded independently, and the best result is selected
according to the maximum likelihood 
criterion, typically based on minimum weight or log-likelihood
score. Since each decoding attempt is independent, they can be
executed in parallel, introducing minimal latency overhead. However,
in the quantum setting, the structure of qLDPC codes introduces unique
advantages. Due to the degeneracy and high distance of qLDPC codes,
and the tendency of BP decoders to favor minimum-weight solutions,
successful BP convergence rarely results in a logical error. This is
partly because the classical codes $\mathcal{C}_X$ and
$\mathcal{C}_Z$, defined by the parity-check matrices $H_X$ and $H_Z$,
have low minimum distances (recall that the minimum distance of
$\mathcal{C}_X$ is upper bounded by the row weight of $H_Z$, and vice
versa). As a result, there are many low-weight codewords in
$\mathcal{C}_X$ and $\mathcal{C}_Z$. Therefore, even if BP converges
to a non-optimal codeword within $\mathcal{C}_X$ or $\mathcal{C}_Z$,
the resulting error is likely to differ from the optimal solution by a
low-weight codeword. Consequently, the probability that this
low-weight codeword forms a logical operator is very low, as the
weight of logical operator is much higher (at least $d$).  This
motivates our use of a speculative decoding strategy. In our BP-SF approach,
we omit the maximum likelihood selection step: Because of code
degeneracy, any solution that satisfies the syndrome is likely to
belong to the correct coset, particularly in the low-error regime and
for high-distance codes. As a result, we simply return the first valid
codeword that satisfies the syndrome among the parallel decoding
attempts. This strategy reduces latency while preserving decoding
performance.

\section{Simulation Results}

In this section, we evaluate the performance of our proposed BP-SF
decoder and compare it against the BP-OSD decoder. All BP decoders use
the min-sum algorithm with an adaptive damping factor $\alpha=1-2^{-i}$,
where $i$ is the current number of iterations. The OSD method is
OSD-CS in~\cite{roffe2020}, and for brevity, we use labels such as
``BP1000-OSD10'' to denote a decoder using BP with a maximum of 1,000
iterations followed by OSD-CS of order 10. Each data point is obtained
by collecting at least 100 logical errors unless specified
otherwise. The statistical uncertainty is sufficiently small, and
error bars are omitted for clarity.

\subsection{Code Capacity Model}
In the code capacity error model, $X$, $Y$, and $Z$ errors are applied
independently for each data qubit with probability of $p/3$. All other
operations are perfect, where $p$ is the physical error rate.  In this
model, we decode all codes using all trial vectors of weight only up
to 1, which proves sufficient to achieve satisfactory logical error
rates.  In many of the codes we tested, the BP decoder already
performs well under the code capacity model, leaving limited room for
improvement as shown in Appendix~\ref{app:capacity}. However, there
exist some exceptions. One such case is the
$\llbracket154, 6, 16\rrbracket$ coprime BB code introduced
in~\cite{wang2024coprime}, where the min-sum decoder performs poorly
despite the code's high distance.

\begin{figure}[htb]
    \centering
    \includegraphics[width=3.1 in]{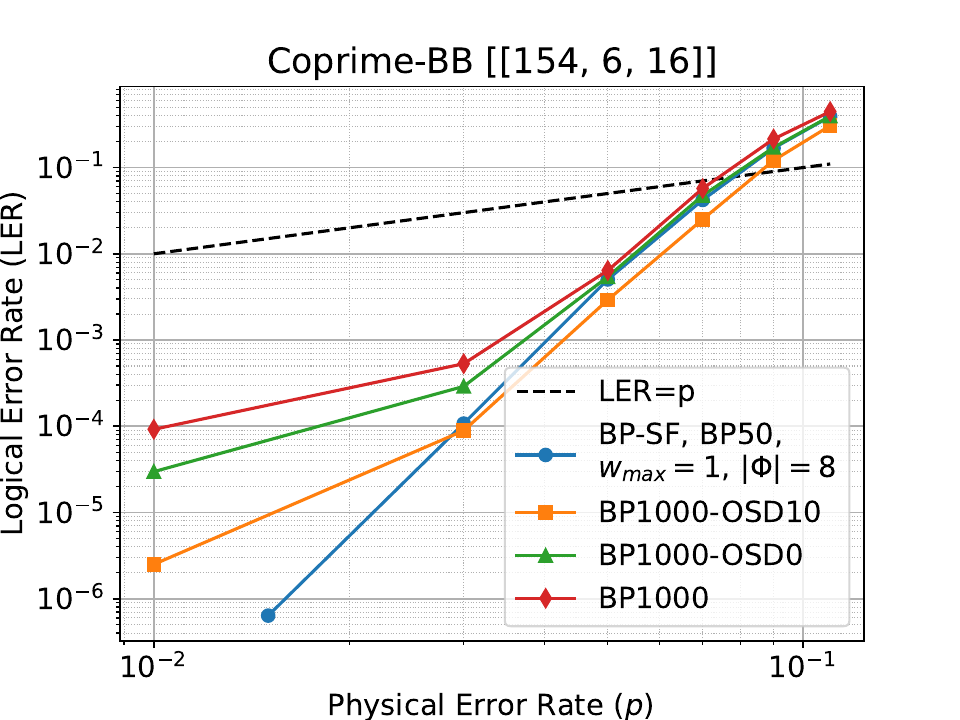}
    \caption{Error rates of the $\llbracket 154, 6, 16\rrbracket$
      coprime-BB code under the code capacity model.}
    \label{154_capa}
\end{figure}

\begin{figure}[htb]
    \centering
    \includegraphics[width=3.1 in]{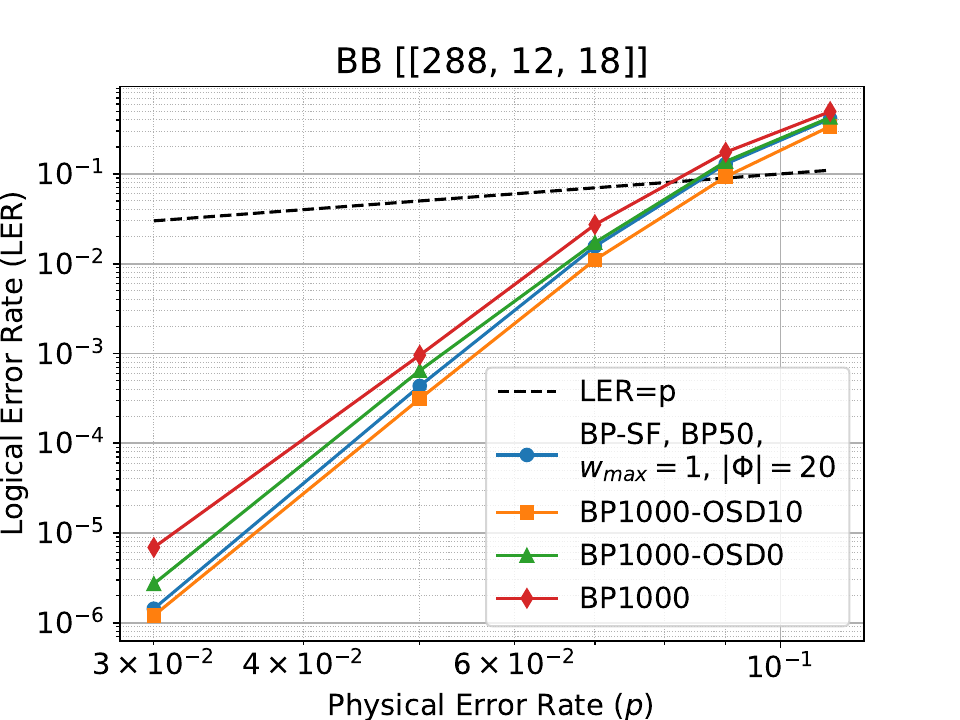}
    \caption{Error rates of the $\llbracket 288, 12, 18\rrbracket$ BB
      code under the code capacity model.}
    \label{288_capa}
\end{figure}

Fig.~\ref{154_capa} shows the logical error rates of different
decoders on the $\llbracket 154, 6, 16\rrbracket$ coprime-BB code
under the code capacity noise model. For our BP-SF decoder, the
candidate set size is set to $|\Phi|=8$, resulting in a maximum of
$50\times(8+1)=450$ BP iterations per decoding attempt. Considering
that BP decoders can be run in parallel after the initial run, the
latency can be optimized to 100 BP iterations. As shown, our BP-SF
decoder significantly outperforms both the baseline BP and BP-OSD
decoders, achieving lower logical error rates with fewer iterations
and without requiring costly OSD post-processing. Additionally, both
BP and BP-OSD exhibit an error floor at low physical error rates. Upon
examining the decoding failures, we find that many are due to
low-weight (e.g., weight-3) errors that fall into trapping sets.  Our
BP-SF algorithm circumvents this issue by decoding flipped
syndromes, which effectively reduces the number of errors in decoding
attempts.
Another example is the $\llbracket 288, 12, 18\rrbracket$ BB code
from~\cite{Bravyi_2024}.  As shown in Fig.~\ref{288_capa}, our
BP-SF decoder performs similarly to the BP-OSD decoder while using
fewer than $50\times(20+1)=1050$ iterations per decoding attempt and
the latency is still 100 iterations considering full parallelization,
as we analyzed above.

\subsection{Circuit-Level Noise Model}

Under the circuit-level noise model, errors are injected uniformly
across gates and measurements. They can propagate through the circuit
as it runs. We use Stim~\cite{gidney2021stim} to generate the syndrome
extraction circuit, the detector error model, and the corresponding
parity-check matrix. In this matrix, each row represents a detector
event, and each column corresponds to a specific error
mechanism. Following the convention in prior literature, we perform
$d$ rounds of syndrome extraction and define the logical error rate
per round as
\begin{equation}
    \text{LER Per Round}= 1- ( 1-\text{LER})^{\frac{1}{d}},
\end{equation}
where $\text{LER}$ is the logical error rate after $d$ rounds. 

In this model, the number of error mechanisms is typically much larger
than the number of qubits, resulting in very large parity-check
matrices. Consequently, flipping a single bit as in the code capacity
model is often insufficient for the BP decoder to converge. On the
other hand, exhaustively decoding all trial vectors with weight up to
a threshold is computationally expensive. To address this, we adopt a
sampling-based approach. Given a maximum trial vector weight
$w_{max}$, we randomly sample $n_s$ trial vectors for each weight in
$\{1,\dots,w_{max}\}$, resulting in a total of $n_s\times w_{max}$
trial vectors per failed BP decoding attempt.

\begin{figure}[htb]
    \centering
    \includegraphics[width=3.2 in]{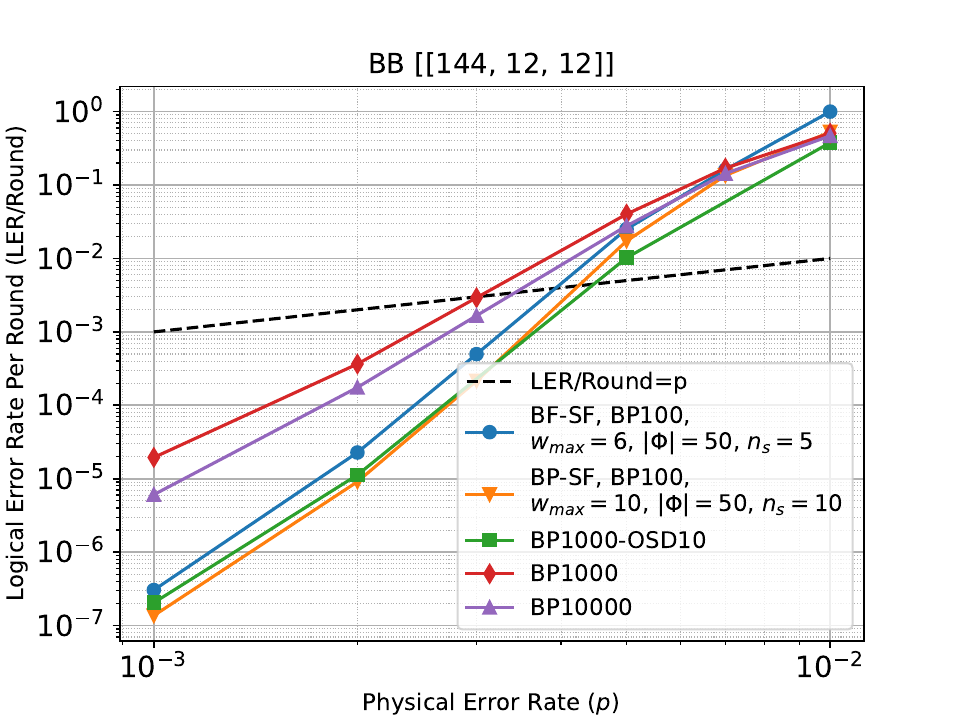}
    \caption{Error rates of the $\llbracket 144, 12, 12\rrbracket$ BB
      code under the circuit-level noise model.}
    \label{144_circ}
\end{figure}

Figure~\ref{144_circ} shows the logical error rates of different
decoders on the $\llbracket 144, 12, 12\rrbracket$ BB code under the
circuit-level noise model. In this setting, the number of error
mechanisms is approximately 8,000, which is significantly larger than
the code length used in the code capacity model. Therefore, we expand
the candidate set size in our BP-SF decoder. Specifically, our
decoder uses up to 3,100 BP iterations (based on
$w_{max}=6, n_s=5$) and achieves logical error rates that are slightly
higher but still comparable to that of the BP-OSD decoder, which uses
a maximum of 1,000 iterations and OSD order 10.

\begin{figure}[htb]
    \centering
    \includegraphics[width=3.1 in]{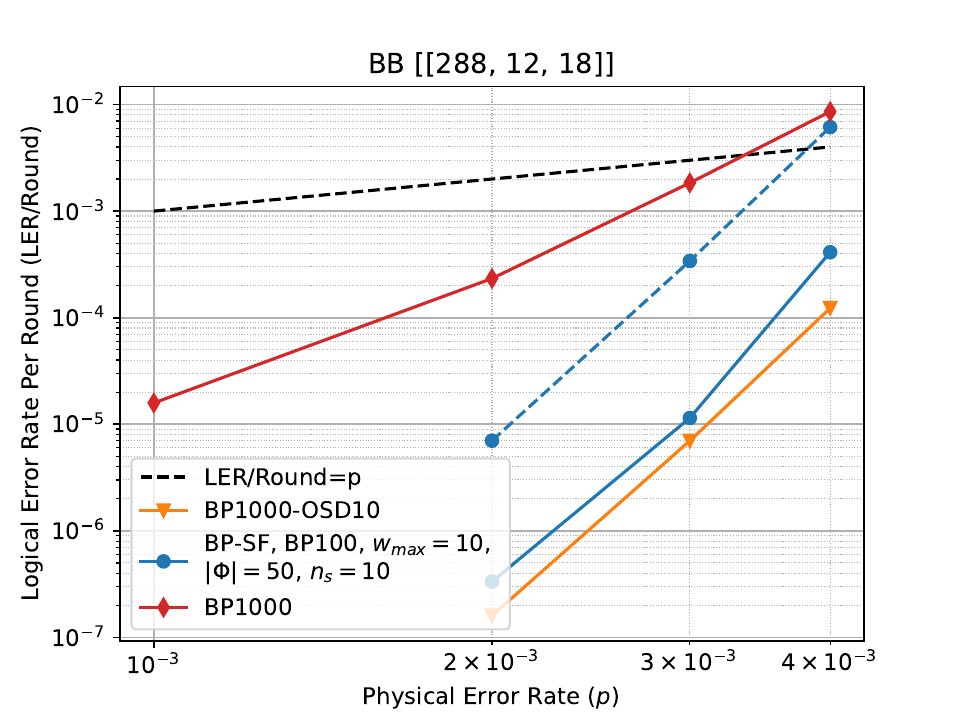}
    \caption{Error rates of the
      $\llbracket 288, 12, 18\rrbracket$ BB code under the
      circuit-level noise model. Layered BP decoding is used for all
      decoders except the one shown with a dashed line, which is BP-SF using flooding BP.}
    \label{288_circ}
\end{figure}

Fig.~\ref{288_circ} shows the logical error rates (LER) of different
decoders on the $\llbracket 288, 12, 18\rrbracket$ BB code under the
circuit-level noise model. The proposed BP-SF decoder exhibits a
slightly higher LER than the BP1000–OSD10 decoder. We note that all
decoders for this code employ the layered BP variant, as the regular
BP shows significantly worse LER performance. This behavior is likely
caused by symmetric trapping sets, as suggested in
\cite{raveendran2021trapping}, which can lead to large variations in
LER depending on the BP scheduling strategy.

\begin{figure}[htb]
    \centering
    \includegraphics[width=3.1 in]{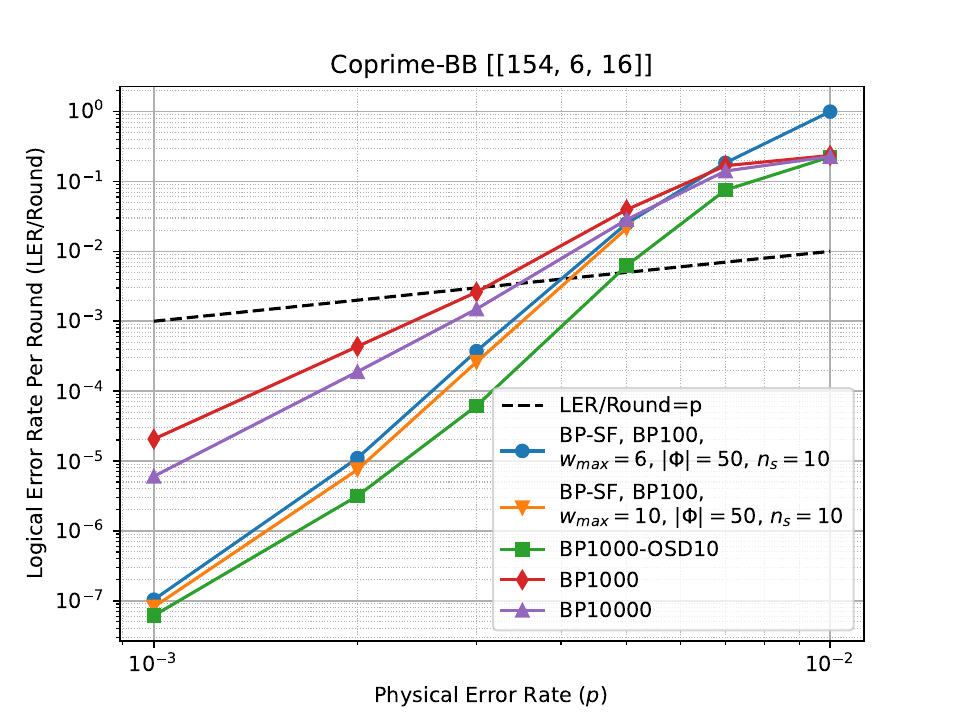}
    \caption{Error rates of the $\llbracket 154, 6, 16\rrbracket$
      coprime-BB code under the circuit-level noise model.}
    \label{156_circ}
\end{figure}

Fig.~\ref{156_circ} shows the logical error rates of different
decoders on the $\llbracket 154, 6, 16\rrbracket$ BB code under the
circuit-level noise model. Our BP-SF decoder uses up to 6,000
iterations (based on $w_{max}=6, n_s=10$). However, since all trial
syndromes are decoded in parallel, the effective decoding latency
corresponds to only 200 iterations if the attempts are run in
parallel. Our BP-SF decoder achieves logical error rates that are
slightly higher but still comparable to that of the BP-OSD decoder at
lower physical error rate, which uses a maximum of 1,000 iterations
and OSD order 10. In the higher physical error rate regime, our BP-SF
decoder exhibits logical error rates that are higher than those of
BP-OSD but still consistently lower than baseline BP decoding.

\begin{figure}[htb]
    \centering
    \includegraphics[width=3.1 in]{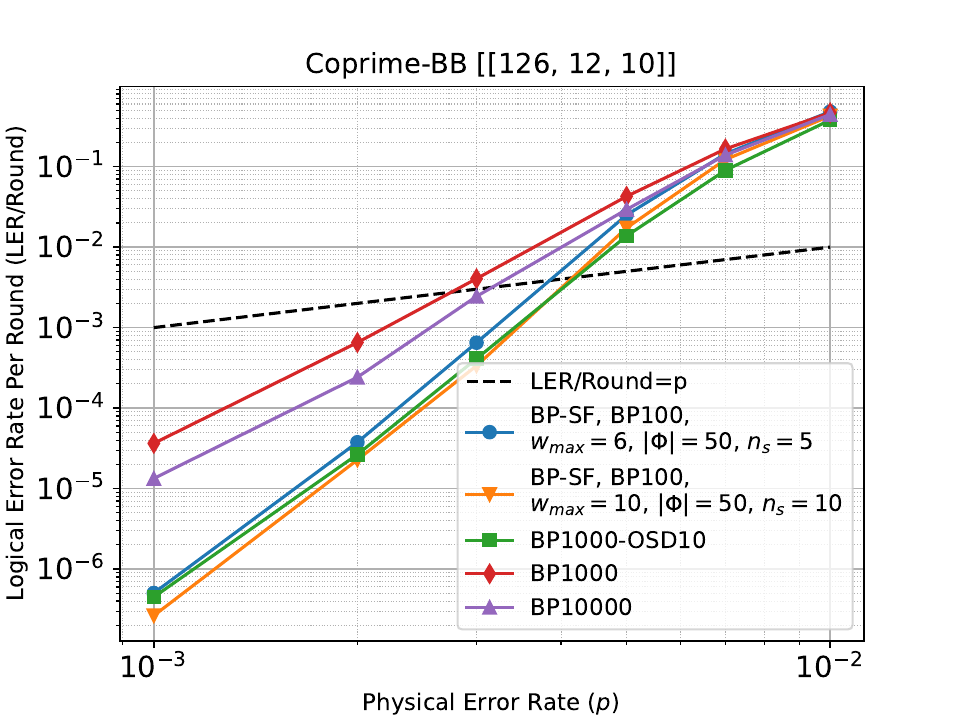}
    \caption{Error rates of the $\llbracket 126, 12, 10\rrbracket$
      coprime-BB code under the circuit-level noise model.}
    \label{126_circ}
\end{figure}

Fig.~\ref{126_circ} shows the logical error rates of different
decoders on the $\llbracket 126, 12, 10\rrbracket$ coprime-BB code
under the circuit-level noise model. Our BP-SF decoder uses up to
approximately 3{,}000 BP iterations to achieve a logical error rate
comparable to that of the BP1000-OSD10 decoder. By increasing both
$n_s$ and $w_{\max}$, we are able to further reduce the logical error
rate to slightly below that of the BP-OSD decoder. However, this
improvement comes at the cost of increased complexity, requiring up to
10{,}000 BP iterations.

\begin{figure}[htb]
    \centering
    \includegraphics[width=3.1 in]{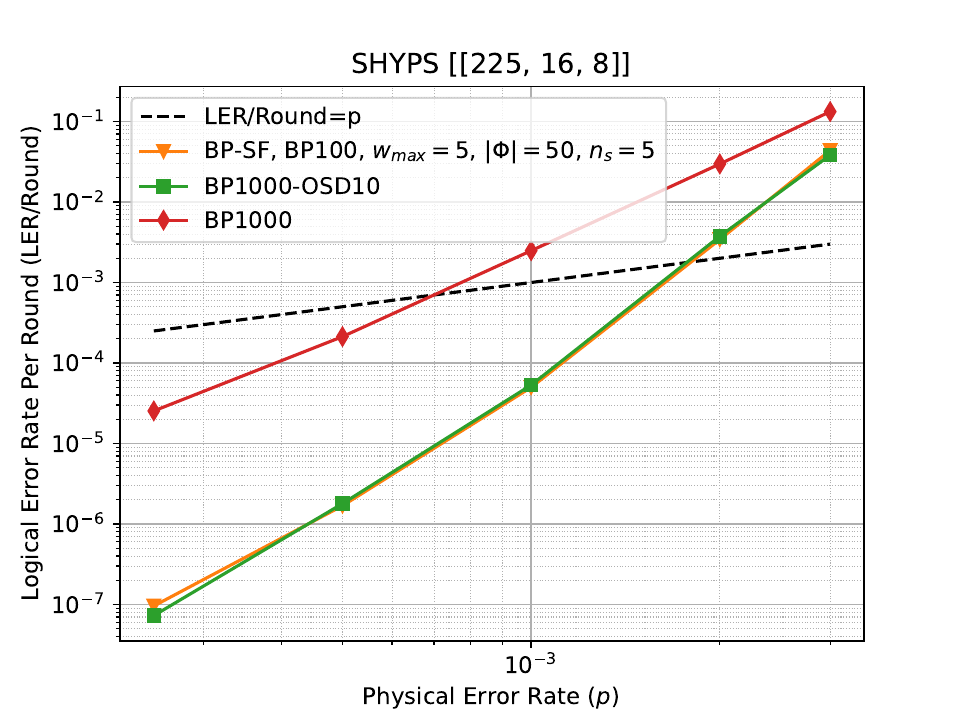}
    \caption{Error rates of the $\llbracket 225, 16, 8\rrbracket$
      SHYPS code under the circuit-level noise model.}
    \label{shyps_circ}
\end{figure}

To demonstrate that our decoder generalizes across different classes
of qLDPC codes, we evaluated BP-SF on the
$\llbracket225,16,8\rrbracket$ subsystem hypergraph product simplex
(SHYPS) code~\cite{malcolm2025computing}, as shown in
Fig.~\ref{shyps_circ}. The proposed BP-SF achieves nearly an identical
logical error rate (LER) performance to the BP1000–OSD10 decoder,
using $w_{\text{max}}=5$ and only $n_s=5$, i.e., with fewer parallel
trials than required for other codes.

\subsection{Complexity and Parameter Selection}

Next, we analyze the computational complexity of our BP-SF decoder. To
ensure a fair comparison with baseline methods, we measure the average
number of BP iterations under a serial execution model. Specifically,
when the initial BP decoding fails, each trial syndrome is decoded
sequentially, and the total number of iterations is defined as the
cumulative number of BP iterations required until the first successful
decoding. This approach provides a conservative estimate of decoding
cost, as it does not account for the inherent parallelism of our
method, but allows for a meaningful comparison with standard decoders.

\begin{figure}[htb]
    \centering
    \includegraphics[width=3.1 in]{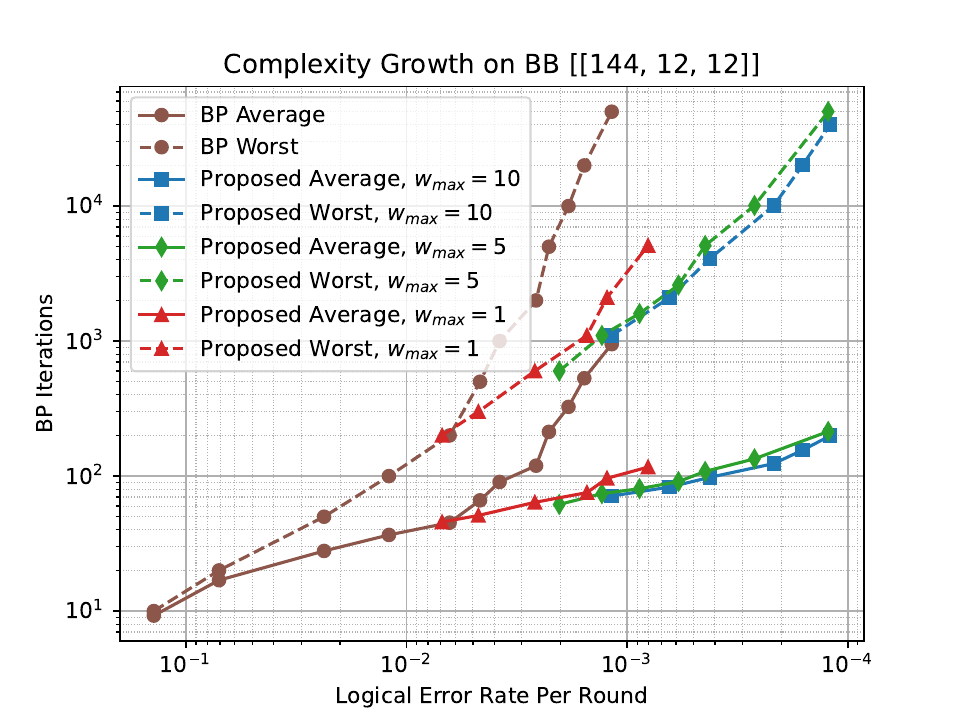}
    \caption{Complexity growth of different decoders. The parameter
      $|\Phi|$ is set to 50 for all BP-SF decoders. The number of
      iterations is calculated assuming serial execution.}
    \label{fig:complexity_growth}
\end{figure}

Fig.~\ref{fig:complexity_growth} shows the growth in decoding
complexity for the $\llbracket 144, 12, 12\rrbracket$ code as we
target progressively lower logical error rates. For the BP decoder, we
vary the maximum number of iterations to control decoding
complexity. For our BP-SF decoder, we fix the maximum number of
iterations per BP instance to 100 and vary $n_s$ (the number of trial
vectors sampled per weight), while keeping $w_{\text{max}}$
constant. This allows us to explore the trade-off between complexity
and performance. The physical error rate is fixed at
$3 \times 10^{-3}$ under the circuit-level noise model. All data
points are collected by simulating 10{,}000 shots for logical error
rates above $10^{-3}$ and 100{,}000 shots for those below.

Across all decoders, we observe a linear region in which the number of
BP iterations increases approximately linearly as the logical error
rate decreases, up to a point where each curve drops off sharply,
forming what we refer to as a cliff. This cliff marks a regime where
the decoder can no longer reliably suppress logical errors within
reasonable iteration limits. Our BP-SF decoder consistently
postpones this cliff compared to baseline BP, maintaining a lower
iteration count at comparable logical error rates. Moreover,
increasing $w_{max}$ increases the complexity but extends the linear
region further and delays the start of the cliff, providing a tunable
trade-off between decoding complexity and error suppression.

\begin{figure}[htb]
    \centering
    \includegraphics[width=3.1 in]{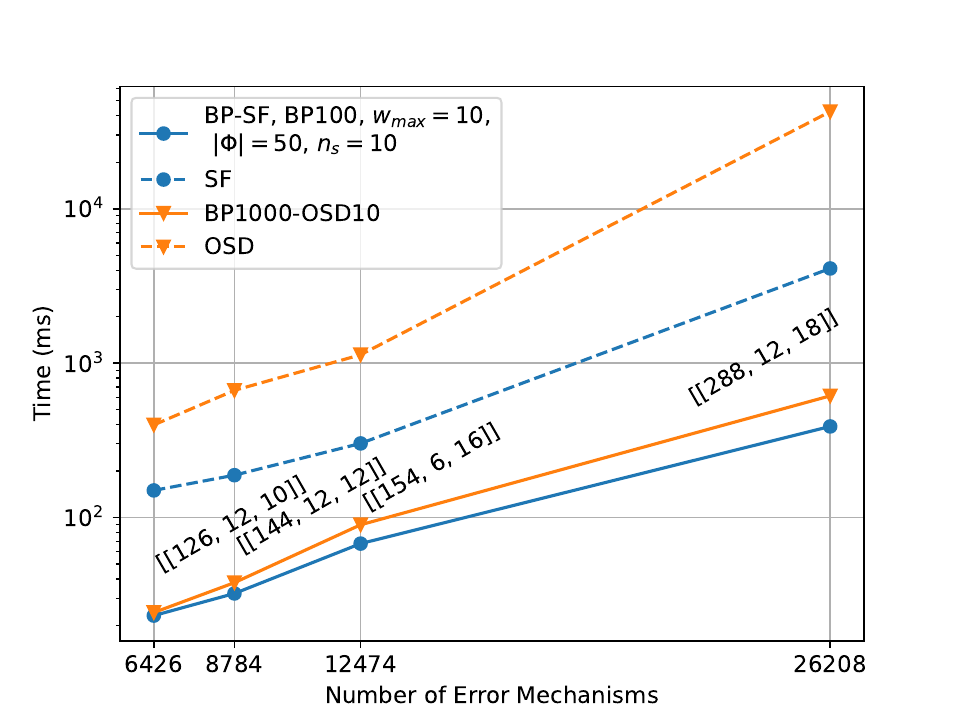}
    \caption{Latency scaling of BP-SF and BP-OSD at a physical error
      rate of $3\times10^{-3}$. Dashed lines indicate the average
      latency of the post-processing stage, measured only for cases
      where the initial BP decoding fails.}
    \label{fig:scale}
\end{figure}

Fig.~\ref{fig:scale} shows how the decoding latency scales as the code
size increases. Here, ``number of error mechanisms'' refers to the total
number of error sources in a memory experiment circuit, which grows
much faster than the number of qubits. The proposed BP-SF consistently
achieves lower average decoding latency. Since most errors are
corrected by the initial BP stage, the asymptotic advantage of BP-SF
is not evident from the average decoding time alone, which is about
$0.63\times$ that of BP-OSD for the $\llbracket288,12,18\rrbracket$
code. However, when considering only cases where the initial BP fails,
BP-SF requires merely $0.1\times$ the latency of BP-OSD, an order of
magnitude improvement. Both the overall average latency and the
post-processing latency of BP-SF can be further reduced (discussed in
Section~\ref{sec:runtime}).

\section{Parallel Implementation and Performance}
\label{sec:runtime}

In this section, we describe the implementation of our proposed
algorithm and compare it with the BP-OSD decoder in the LDPC
library~\cite{Roffe_LDPC_Python_tools_2022} and Nvidia's CUDA-Q
library~\cite{cudaq}. For a fair comparison, we implemented three
versions of our algorithm:

\textbf{Serial CPU version:} We modify the BP decoder
from~\cite{Roffe_LDPC_Python_tools_2022}, which is also a serial CPU
version implemented in C++ and wrapped with Cython, to track bit-level
oscillations. If the first decoding attempt fails, trial vectors and
corresponding syndromes will be prepared in Python, and each trial
syndrome will be decoded using the C++ BP decoder. We return the first
successful decoding result and skip the rest.

\textbf{Parallel CPU version (P=\# of worker processes):}
This version maintains a persistent worker process pool along with
input and output queues to handle the remaining $n_s \times w_{\max}$
decoding trials after the initial BP attempt fails. The manager
process selects candidate positions, generates trial vectors, and
computes the corresponding trial syndromes. These trial syndromes are
then split into small batches and placed in the input queue. Worker
processes continuously retrieve batches from the queue and attempt to
decode them until one finds a valid solution, which it places in the
output queue. The main process monitors the output queue and, once a
valid result is found, signals all workers to stop. The workers then
wait for the next input from the input queue, and the main process returns
the successful result. To avoid accepting stale results, each syndrome
is tagged with a serial number. The main process compares the serial
number of each fetched result with that of the currently processed
syndrome to ensure correctness.

\textbf{Estimated GPU version (GPU\_Est):}
Since the CUDA-Q library does not provide access to oscillation
statistics during decoding, we implemented this version to estimate
the decoding speed. We first precompute the syndromes and their
corresponding oscillation bits using a CPU-based BP decoder. These
syndromes are then decoded using CUDA-Q. If decoding fails, the
precomputed oscillation bits are used in Python to generate trial
syndromes, which are then decoded one-by-one using CUDA-Q until a
successful decoding is found or the maximum number of trials is
reached. Due to numerical instability and implementation differences,
the CUDA-Q decoder may report a different success status than the CPU
BP decoder. In such cases, the pre-computation is invalid and we
discard those results. This approach results in a pessimistic estimate
of GPU performance, as trial syndrome generation is performed on the
CPU after initial failure, and transferring them to the GPU introduces
additional memory copy overhead. A more efficient strategy would
submit all trial syndromes to the GPU decoder simultaneously and
return upon the first success. However, the current CUDA-Q library
only supports the \texttt{decode\_batch} method, which waits for all
syndromes in the batch to complete, effectively blocking on the
slowest one.

For the sake of fairness, we select the number of BP iterations to
make BP-OSD as fast as possible. At first glance, one might expect
that reducing the number of BP iterations would lower the overall
decoding latency. However, beyond some point, reducing the number of
BP iterations not only degrades the logical error rate but can also
\emph{increase} the total latency as shown in
Table~\ref{tab:bposd_selection}. This is caused by the BP stage being
relatively inexpensive compared to the OSD stage, where a higher number
of BP iterations allows BP to correct more errors and reduces the
frequency with which the costly OSD procedure must be invoked. We
listed the BP iterations of experiments with the
$\llbracket144,12,12\rrbracket$ code under circuit-level noise of
$p=3\times 10^{-3}$.

\begin{table}[htb]
    \centering
    \begin{tabular}{|c|c|c|}
    \hline
         Decoder&  LER/$d$ @ 3e-3 & Avg Time @ 3e-3\\
         \hline
         BP100-OSD10 & 2.89 $\times 10^{-4}$&56.13 ms\\
         BP400-OSD10& 2.23 $\times 10^{-4}$  &37.69 ms\\
         BP1000-OSD10& 2.11 $\times 10^{-4}$ &36.44 ms\\
         BP2000-OSD10&  2.00 $\times 10^{-4}$ &44.01 ms\\
         BP10000-OSD10& 1.84 $\times 10^{-4}$ &94.94 ms\\
         \hline
    \end{tabular}
    \caption{Logical error rate per round and average decoding
      time for BP-OSD with different iterations.}
    \label{tab:bposd_selection}
\end{table}

We note that for CPU-based decoders, it is possible to parallelize
decoding using OpenMP within each BP iteration or by decoding a batch
of input syndromes in parallel. However, we did not pursue these
options because the BP/BP-OSD decoder provided by the LDPC
library~\cite{Roffe_LDPC_Python_tools_2022} does not currently support
OpenMP acceleration. For fairness and consistency, we instead
implement parallelization at a coarser granularity, i.e., by decoding
trial syndromes in parallel. This approach may be
less efficient than parallelizing BP iterations directly, as it only
accelerates cases where the initial decoding attempt fails. However,
it helps mitigate the long-tail latency caused by the large number of
trials required in such cases. We also do not parallelize across input
syndromes, as decoding them sequentially is more aligned with
real-world use cases, where syndrome extraction is performed
sequentially and syndromes arrive in a streaming fashion.

Benchmarks were conducted on an NVIDIA Tesla V100-SXM2-16GB GPU and an
Intel Xeon E5-2698 v4 @ 2.20GHz CPU. For each physical error rate,
20{,}000 syndromes were tested. Our newly proposed decoder with 100 BP
iterations, $w_{\max} = 10$, and $n_s = 10$ achieves nearly identical
logical error rates to BP1000-OSD10 ($1,000$ BP and OSD-CS of order $10$)
below threshold for the $\llbracket 144,12,12 \rrbracket$
code, which allows us to adopt these settings for all evaluations in
this section.

\begin{figure}[htb]
    \centering
    \includegraphics[width=3.3 in]{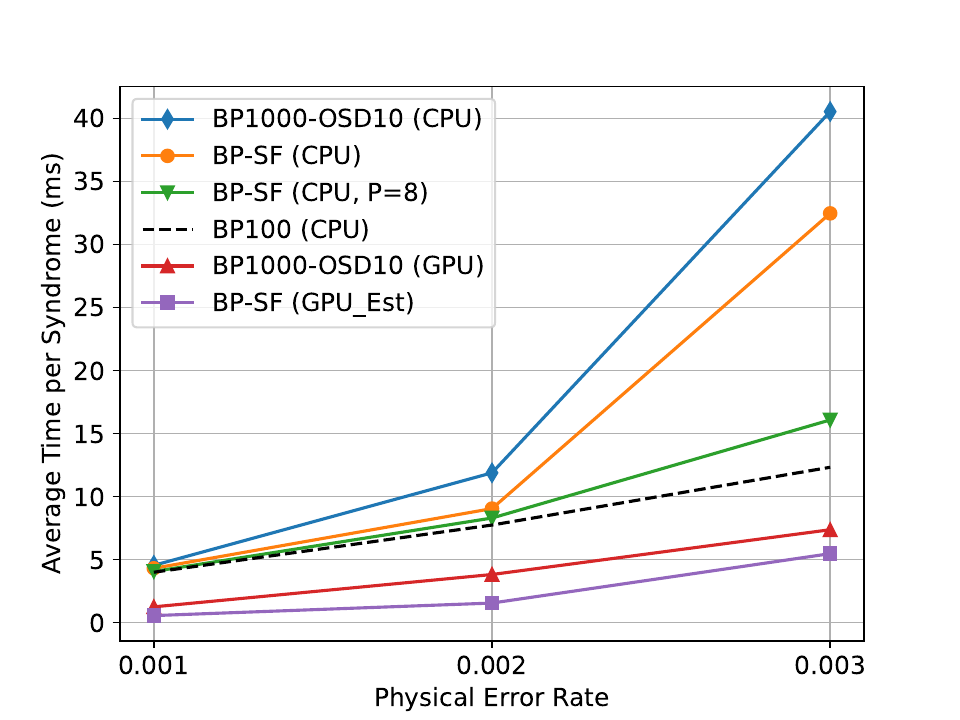}
    \caption{Average decoding time per syndrome for the BP-SF decoders
      compared with BP1000-OSD10 and BP100, under varying physical
      error rates. BP100 is performed on the CPU serially as a
      reference lower bound, as it performs no post-processing.}

    \label{run_time_plot}
\end{figure}

Figure~\ref{run_time_plot} shows the average decoding time per
syndrome under varying physical error rates. The BP-OSD implementations 
for CPU and GPU are obtained from the LDPC Python library and NVIDIA CUDA-Q, respectively.
The GPU versions consistently achieve the lowest average runtime due
to their high degree of parallelism. At low physical error rates
(e.g., 0.001), the BP-SF decoders and BP1000-OSD10 exhibit similar
performance, as most syndromes are successfully decoded during the
initial BP attempt without invoking post-processing. However, as the
error rate increases, the BP-SF decoder outperforms BP1000-OSD10 thanks to
its more efficient post-processing strategy.
The parallel CPU version (CPU, P=8) achieves about 1.8$\times$ speedup
over the serial version and approaches the performance of BP100, which
is included as a lower bound since it performs no post-processing.
The GPU runtime of decoders increases only slightly with the error
rate, due to the ability to perform computations in parallel. We can
also see that the BP1000-OSD10 from CUDA-Q is
slightly slower than our GPU\_Est on the same platform.

\begin{figure}[htb]
    \centering
    \includegraphics[width=3.1 in]{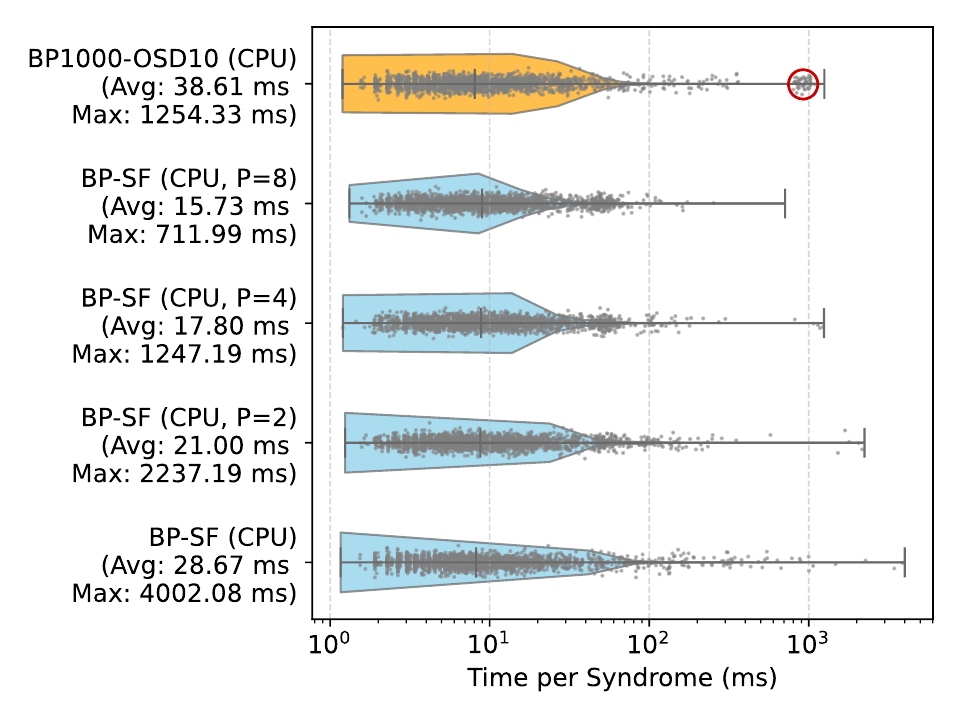}
    \caption{Distribution of single-syndrome decoding times for
      different decoder configurations when physical error rate is
      0.003. The horizontal violin plots show the probability density
      of the time taken for each decoding attempt, plotted on a
      logarithmic scale. Overlaid gray points represent individual
      decoding events, with vertical jitter added for clarity. The
      vertical lines mark the min, median, and max decoding time.
      Decoding events invoking OSD stage are circled in red.}

    \label{time_dist_cpu}
\end{figure}

Figure~\ref{time_dist_cpu} shows the distribution of decoding times for
different decoders at a physical error rate of 0.003. All versions of
our BP-SF decoder exhibit lower average decoding times compared to
BP1000-OSD10 (Avg: 38.61 ms). Notably, the BP1000-OSD10 curve shows a
distinct gap, corresponding to cases where OSD post-processing is
triggered. In contrast, the serial CPU version of our algorithm
displays a long-tailed but more compact distribution, due to most syndromes
that fail the initial decoding being resolved within a few additional
trials, with only a small number of outliers requiring longer decoding
times. As the number of parallel processes increases, this long tail
becomes increasingly compressed, resulting in even lower average
decoding times, e.g., 21.00 ms with 2 processes, 17.80 ms with 4, and
15.73 ms with 8. And the speedup in worst case for P$=$8 is
5.6$\times$ compared with the serial version. This demonstrates the
efficiency of our approach.

\begin{figure}[htb]
    \centering
    \includegraphics[width=3.0 in]{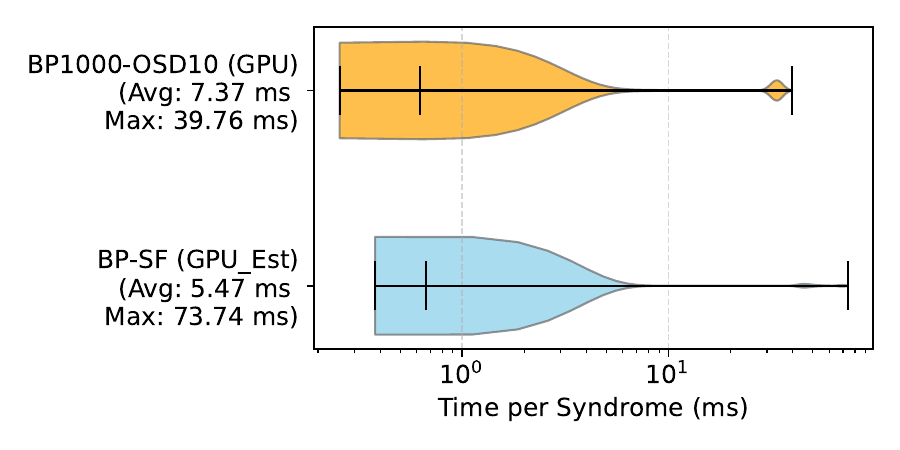}
    \caption{Distribution of single-syndrome decoding times for
      GPU decoders when physical error rate is
      0.003.  }

    \label{time_dist_gpu}
\end{figure}

Figure~\ref{time_dist_gpu} shows the runtime distribution for the GPU 
decoders. We observe a similar pattern: while the BP-SF method achieves 
a lower average runtime (5.47 ms vs. 7.37 ms), its maximum runtime 
(73.74 ms) is higher than that of the BP-OSD decoder (39.76 ms), due to the 
serial decoding of trial syndromes. This long-tail latency can be effectively 
mitigated by enabling a GPU function that accepts a batch of trial syndromes 
and returns as soon as any one is successfully decoded as GPU can naturally 
decode multiple syndromes in parallel. We also observe that the BP-SF 
decoder exhibits a slightly higher minimum runtime (approximately 0.1 ms), 
which may be attributed to the  use of a wrapper function around the CUDA-Q 
decoder, which introduces additional parameter I/O and initialization overhead.

\textbf{Discussion:} In addition to CPU and GPU implementations, our
BP-SF decoder is also well-suited for deployment on FPGAs and ASICs
due to its reliance on components that are already widely adopted in
hardware. The core of the decoder is based on BP, which has mature
hardware implementations in commercial systems such as 5G mobile
devices~\cite{su202258}, 10GBASE-T Ethernet~\cite{5437474}, digital
video broadcasting~\cite{1657127}, etc. The remaining components, such
as candidate selection and trial syndrome generation, are also
lightweight and hardware-friendly. Candidate selection involves
identifying the top $|\Phi|$ most oscillating bits, which can be
efficiently implemented using partial sorting algorithms such as
incomplete selection sort, or full quicksort for larger values of
$|\Phi|$. Trial syndrome generation can be formulated as a sparse
matrix-sparse vector (SpMSpV) multiplication, another operation that maps well to
hardware accelerators. Together, these features make our decoder
highly amenable to low-latency, energy-efficient hardware realization.
In a superconducting based quantum computer, assuming a typical
syndrome extraction round time on the order of 1 $\mu$s and a BP
iteration latency of approximately 20 ns~\cite{9562513}, our decoder
can achieve a worst-case latency of about 4 $\mu$s when fully
parallelized (corresponding to 200 BP iterations, 100 for the initial
BP stage and 100 for the parallelized trials). Considering that a full
syndrome extraction circuit typically requires $d$ rounds of syndrome
extraction, our decoder is fast enough to perform real-time decoding.

\section{Conclusion and Future Work}
\label{sec:conc}

We introduced a fully parallelizable decoder based on belief
propagation (BP). By leveraging speculative decoding and bit-flipping
strategies guided by BP oscillation statistics, our proposed BP-SF
method achieves logical error rates comparable to BP-OSD, while
significantly reducing computational complexity and avoiding costly
Gaussian elimination. Extensive simulations show that our decoder
performs exceptionally well under the code-capacity noise model across
a range of bivariate bicycle codes. Under the more realistic
circuit-level noise model, the decoder still delivers reasonable
performance, though it requires a larger number of decoding trials to
achieve comparable accuracy. 
We also note that
throughout the paper, we adopt the widely used min-sum variant of BP
because of its simplicity and computational efficiency. Nonetheless,
our approach could potentially benefit from incorporating more
advanced BP-based techniques as long as their convergence is also
affected by oscillating bits.

In future work, we aim to better understand the challenges posed by
circuit-level noise and explore targeted improvements, such as more
effective candidate selection, improved trial vector sampling
strategies, efficient decoder implementation, and enhancements to the
inner BP decoder, in order to further improve decoding performance in
practical fault-tolerant quantum computing systems.

\section*{Acknowledgment}
The authors would like to thank John Stack for providing source code
and for discussions on circuit-level noise simulation, and Timo
Hillmann for valuable discussions regarding the BP-OSD decoder.
This material is based upon work supported by the U.S. Department of
Energy, Office of Science, National Quantum Information Science
Research Centers, Co-design Center for Quantum Advantage (C2QA) under
contract number DE-SC0012704, (Basic Energy Sciences, PNNL FWP
76274). This research used resources of the Oak Ridge Leadership
Computing Facility, which is a DOE Office of Science User Facility
supported under Contract DE-AC05-00OR22725. This research used
resources of the National Energy Research Scientific Computing Center
(NERSC), a U.S. Department of Energy Office of Science User Facility
located at Lawrence Berkeley National Laboratory, operated under
Contract No. DE-AC02-05CH11231. The Pacific Northwest National
Laboratory is operated by Battelle for the U.S. Department of Energy
under Contract DE-AC05-76RL01830.
This work was also supported in part by NSF awards
OSI-2410675, PHY-1818914, PHY-2325080, MPS-2120757, CISE-2217020, and
CISE-2316201 as well as DOE DE-SC0025384.

\appendix
\subsection{Code Construction}

\begin{figure*}[h]
\label{app_capa}
    \label{good_capa}
     \centering
    \begin{subfigure}{0.32\textwidth}
  \includegraphics[width=\textwidth]{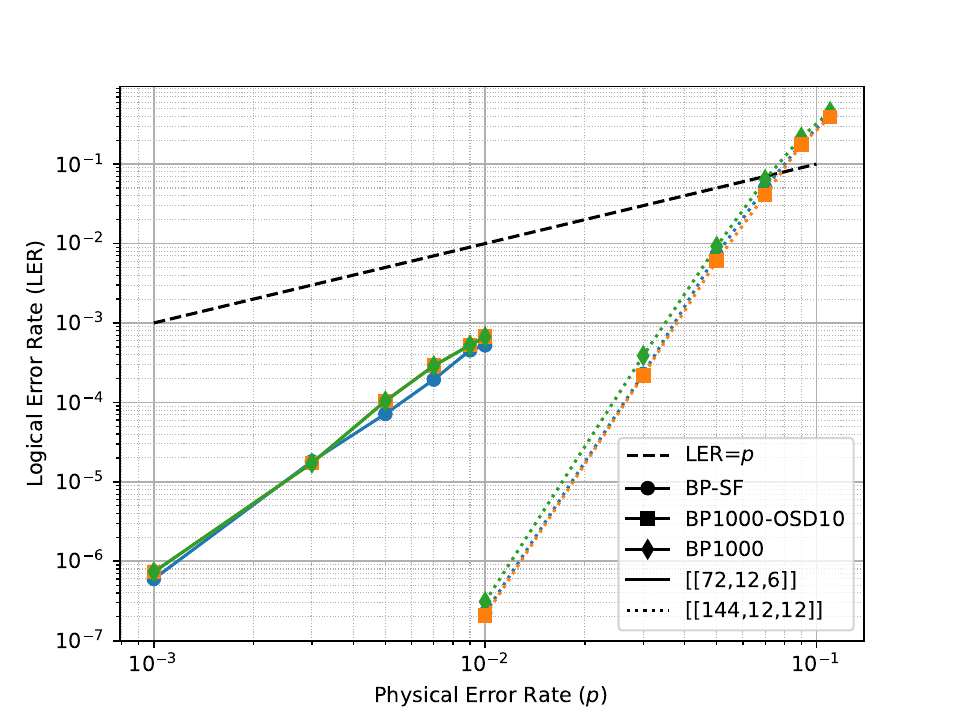}
  \caption{}
  \end{subfigure}
  \hfill
  \begin{subfigure}{0.32\textwidth}
  \includegraphics[width=\textwidth]{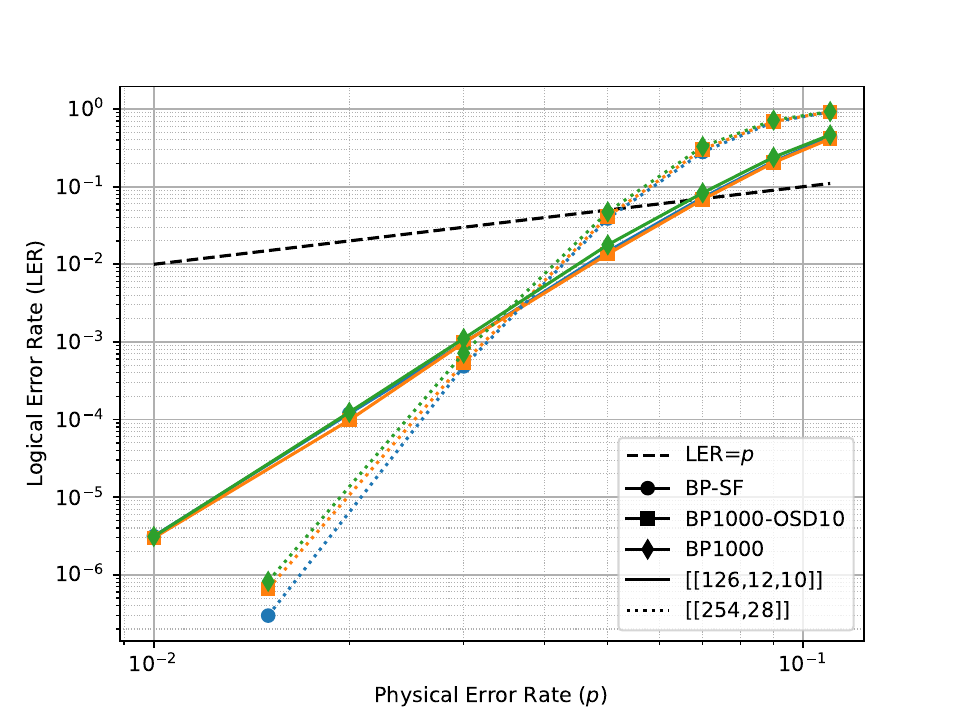}
    \caption{}
    \end{subfigure}
    \begin{subfigure}{0.32\textwidth}
    \includegraphics[width=\textwidth]{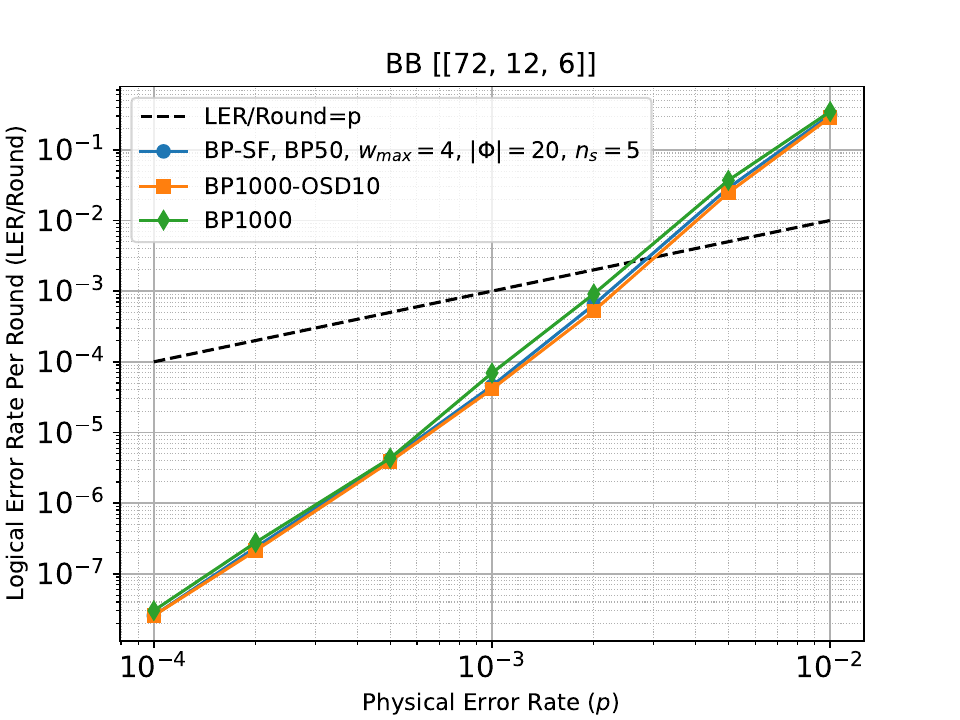}
    \caption{}
    \label{72_circ}
  \end{subfigure} 

  \caption{(a-b): The logical error rates of different codes under code
    capacity error model. For the $\llbracket72,12,6\rrbracket$ BB
    code, BP and BP-OSD have the same error when we set the seed to
    the same. BP-SF has $w_{max}=1$, and  $|\Phi|=4,7,6,13$ for $\llbracket72,12,6\rrbracket,\llbracket144,12,12\rrbracket,\llbracket126,12,10\rrbracket,\llbracket254,28\rrbracket$ codes, respectively. (c): Error rate comparison on the $\llbracket72,12,6\rrbracket$ under circuit-level noise model.}
\label{fig:decode}
\end{figure*}
In this section, we provide the construction parameters for the qLDPC
codes referenced throughout the paper. The construction and
circuit-level implementation of SHYPS code can be found
in~\cite{malcolm2025computing}.
\subsubsection{GB codes}
Let $S_l$ be the shift matrix of size $l$, defined as
\begin{equation}
    S_l=I_l >> 1,
\end{equation}
where ``$>>$'' denotes the right cyclic shift for each row in the
matrix, and $I_l$ is the identity matrix of size $l$. For example,
\begin{equation}
    S_3 = I_3>> 1 = \begin{bmatrix}
        0&1&0\\
        0&0&1\\
        1&0&0
    \end{bmatrix}.
\end{equation}
Let $x=S_l$ here. The GB code can be defined by two polynomials,
$a(x)$ and $ b(x)$. These two polynomials can be represented by two
matrices $A, B$ naturally, and the parity check matrices for the GB
code are defined as
\begin{equation}
    \begin{split}
        H_X&=[A|B]\\
        H_Z&=[B^T|A^T].
    \end{split}
\end{equation}

The $\llbracket 254,28\rrbracket$ GB code used in this paper can be
constructed by $l=127$, $a(x) = 1+x^{15}+x^{20}+x^{28}+x^{66}$,
$b(x)=1+x^{58}+x^{59}+x^{100}+x^{121}$ as proposed
in~\cite{panteleev2021degenerate}.

\subsubsection{BB Codes}

BB codes are constructed similarly to GB codes but with two
variables. Let $x=S_l\otimes I_m$ and $y=I_l\otimes S_m$, where
$\otimes$ denotes the Kronecker product, the BB codes can also be defined
by two polynomials, $A = a(x,y)$ and $B = b(x,y)$. The parity check
matrices for BB codes are defined similarly to those in GB codes. The
BB codes we used in the paper were proposed in~\cite{Bravyi_2024}. And
the polynomials are shown in Table~\ref{bbcodes}.
\begin{table}[H]
\centering
\caption{BB Codes Used in Simulations}
\label{bbcodes}
\begin{tabular}{|c|c|c|c|c|}
    \hline
    $l$ & $m$ & $a(x,y)$ & $b(x,y)$ & $\llbracket n,k,d\rrbracket $ \\ \hline
    6 & 6 & $x^3+y+y^2$  & $y^3+x+x^2$ &  $\llbracket 72, 12, 6\rrbracket $ \\ \hline 
12 & 6 & $x^3+y+y^2$  & $y^3+x+x^2$ &  $\llbracket 144, 12,12\rrbracket $ \\ \hline
12 & 12 & $x^3+y^2+y^7$  & $y^3+x+x^2$ &  $\llbracket 288, 12,18\rrbracket $ \\ \hline
\end{tabular}
\end{table}

\subsubsection{Coprime-BB Codes}

The coprime-BB code we used originates from~\cite{wang2024coprime}. Let
$\pi=xy$, where $x$ and $y$ are defined the same as in BB codes. The
coprime-BB code we tested can be constructed as shown in
Table~\ref{cbbcodes}.

\begin{table}[H]
\centering
\caption{Coprime-BB Codes Used in Simulations}
\label{cbbcodes}
\begin{tabular}{|c|c|c|c|c|}
    \hline
    $l$ & $m$ & $a(\pi)$ & $b(\pi)$ & $\llbracket n,k,d\rrbracket $ \\ \hline
    7 & 9 & $1+\pi+\pi^{58}$  & $1+\pi^{13}+\pi^{41}$ &  $\llbracket 126, 12, 10\rrbracket $ \\ \hline 
7 & 11 & $1+\pi+\pi^{31}$  & $1+\pi^{19}+\pi^{53}$ &  $\llbracket 154, 6,16\rrbracket $ \\ \hline
\end{tabular}
\end{table}

\subsection{``Good'' Codes for BP}

This subsection presents several codes that demonstrate good performance
under BP decoding, along with their corresponding logical error
rates. Since both BP-OSD and the proposed BP-SF decoder act as
post-processing techniques, they are only invoked when the BP decoder
fails to converge. As a result, for the codes shown below, where BP
alone achieves high success rates, we observe similar overall
performance across all decoding strategies.

\label{app:capacity}
Fig.~\ref{fig:decode}(a-c) shows the performance of various decoders
on several codes under the code-capacity and circuit level noise model. In these cases,
the baseline BP decoder already achieves logical error rates
comparable to those of the BP-OSD decoder, and
post-processing yields only marginal improvements.

\appendix
\section{Artifact Appendix}

\subsection{Abstract}

This artifact contains the source code for the BP-SF decoder, a parallelized belief-propagation decoder for quantum LDPC codes, as described in the paper ``Fully Parallelized BP Decoding for Quantum LDPC Codes Can Outperform BP-OSD''. It includes a custom Cython-based implementation of the Belief Propagation (BP) decoder, scripts for running circuit-level memory experiments using \texttt{stim}, and benchmarking tools to compare performance against standard BP, BP-OSD, and CUDA-Q qLDPC decoders. The artifact allows researchers to reproduce the Logical Error Rate (LER) and decoding speed results presented in the paper.

\subsection{Artifact check-list (meta-information)}
{\small
\begin{itemize}
  \item {\bf Algorithm: }The code repository contains BP-SF, BP-OSD, Standard BP (Min-sum) algorithms.
  \item {\bf Program: }Python 3.11 scripts with Cython extensions.
  \item {\bf Compilation: }C/C++ compiler required for Cython extension.
  \item {\bf Data set: }Synthetic data generated on-the-fly using \texttt{stim} circuit simulations.
  \item {\bf Run-time environment: }Linux; Python 3.11-13 (newer versions may work but are untested)
  \item {\bf Hardware: }Standard CPU for functional tests. NVIDIA GPU with CUDA support recommended for \texttt{cudaq-qec} comparisons and GPU estimation experiments.

  \item {\bf Metrics: } Logical Error Rate per Round (LER/Round), average decoding time per sample (ms).
  \item {\bf Output: }Text logs in \texttt{data/} directory and Jupyter Notebook for plotting.
  \item {\bf How much disk space required (approximately)?: }$\sim$ 200 KBytes.
  \item {\bf How much time is needed to prepare workflow (approximately)?: }$\sim$10 minutes (install dependencies).
  \item {\bf How much time is needed to complete experiments (approximately)?: }Days on a 16-core machine.
  \item {\bf Publicly available?: Yes. \url{https://github.com/Dies-Irae/BP-SF}.
}
  \item {\bf Code licenses (if publicly available)?: }MIT license.
  \item {\bf Data licenses (if publicly available)?: }Not applicable.
  \item {\bf Workflow automation framework used?: }Not applicable.
  \item {\bf Archived (provide DOI)?: }Not available
\end{itemize}
}

\subsection{Description}

\subsubsection{How to access}

The code repository is available at \url{https://github.com/Dies-Irae/BP-SF}.

\subsubsection{Hardware dependencies}
The core BP-SF decoder runs on standard CPUs. To reproduce the GPU comparison baselines (using \texttt{cudaq-qec}), an NVIDIA GPU with CUDA support is required.
\subsubsection{Software dependencies}
\begin{itemize}
    \item OS: Linux.
    \item Python: Version 3.11.
    \item Libraries: \texttt{numpy}, \texttt{scipy}, \texttt{stim}, \texttt{cython}, \texttt{setuptools}, \texttt{matplotlib}.
    \item Optional (for baselines): \texttt{ldpc}, \texttt{cudaq-qec}.
\end{itemize}

\subsection{Installation}
\begin{enumerate}
    \item Clone the repository:
{\small
\begin{verbatim}
git clone git@github.com:Dies-Irae/BP-SF.git
cd BP-SF
\end{verbatim}
}
\item Create and activate a virtual environment:
\begin{verbatim}
python -m venv .venv
source .venv/bin/activate
\end{verbatim}
\item Install Python dependencies:
{\small
\begin{verbatim}
pip install cython numpy scipy 
pip install stim setuptools matplotlib
# Optional: For baselines
pip install cudaq-qec ldpc
\end{verbatim}
}
\item  Build the local Cython extension:
{\small
\begin{verbatim}
export PYTHONPATH=./src_python:$PYTHONPATH
cd minimal_bp_decoder
python setup.py build_ext --inplace
cd ..
\end{verbatim}
}
\end{enumerate}

\subsection{Experiment workflow}
The artifact provides shell scripts to automate the experiments. Results are saved to the \texttt{data/} directory.

\subsubsection{Functional Tests (LER)}
To evaluate the logical error rate of the BP-SF decoder compared to BP-OSD:

{\small
\begin{verbatim}
mkdir data
# Run BP-SF circuit-level simulation
sh bpsf_circ.sh
# Run BP-OSD circuit-level simulation 
# (requires ldpc)
sh bposd_circ.sh
\end{verbatim}
}
\subsubsection{Performance Benchmarks (Timing)}
To measure and compare decoding speeds between BP-SF, BP, BP-OSD, and CUDAQ:

{\small
\begin{verbatim}
# Run speed benchmarks
sh time.sh
\end{verbatim}
}
\emph{Note: The \texttt{time.sh} script runs benchmarks for varying physical error rates and decoders. Ensure \texttt{cudaq-qec} is installed for the GPU baseline.}

\subsection{Evaluation and expected results}

After running the experiments, the results will be stored as text files in the \texttt{data/} folder (e.g., \texttt{time\_cudaq.txt}, \texttt{*bb\_test\_*.txt}).

\subsubsection{Plotting}
Open the Jupyter notebook \texttt{plots.ipynb} and execute the cells to visualize the results.
\begin{itemize}
    \item \textbf{LER Plots:} Comparison of Logical Error Rate vs. Physical Error Rate for BP-SF and BP-OSD (corresponding to Fig. 7-10 in the paper).
    \item \textbf{Timing Plots:} Comparison of average decoding time per sample (corresponding to Fig. 14 in the paper).
\end{itemize}

\bibliographystyle{IEEEtranS}
\bibliography{main}
\end{document}